\documentclass[fleqn,usenatbib]{mnras}

\usepackage{newtxtext,newtxmath}

\usepackage{soul}

\usepackage[T1]{fontenc}
\usepackage{ae,aecompl}
\usepackage{color, colortbl}
\usepackage{wrapfig}
\usepackage{multicol}

\usepackage{graphicx}	
\usepackage{amsmath}	
\usepackage[dvipsnames]{xcolor}

\newcommand{\kms}{\mbox{$\>{\rm km\, s^{-1}}$}}
\newcommand{\pc}{\>{\rm pc}}
\newcommand{\kpc}{\mbox{$\>{\rm kpc}$}}

\newcommand\degrees{^\circ}

\newcommand{\rbr}{\mbox{$r_{\rm br}$}}
\newcommand{\fbr}{\mbox{$f_{\rm br}$}}

\def\ie{{\it i.e.}}

\title{The role of density breaks in driving spiral structure in disc galaxies}

\author[Fiteni et al.]{
Karl Fiteni$^{1}$\thanks{E-mail: karl.fiteni.12@um.edu.mt},
Sven De Rijcke$^{2}$\thanks{E-mail: sven.derijcke@ugent.be},
Victor P.  Debattista$^{1,3}$ and
Joseph Caruana$^{1,4}$
\\
\\
$^{1}$Institute of Space Sciences \& Astronomy, University of Malta, Msida MSD 2080, Malta\\
$^{2}$Ghent University, Dept. Physics \& Astronomy, Krijgslaan 281, S9, B-9000, Ghent, Belgium\\
$^{3}$Jeremiah Horrocks Institute, University of Central Lancashire, Preston PR1 2HE, UK\\
$^{4}$Department of Physics, University of Malta, Msida MSD 2080, Malta
}

\date{Accepted XXX. Received YYY; in original form ZZZ}

\pubyear{2024}

\begin{document}
\label{firstpage}
\pagerange{\pageref{firstpage}--\pageref{lastpage}}
\maketitle

\begin{abstract}
It is well established that stellar discs are destabilized by sharp features in their phase space, driving recurrent spiral modes. We explore the extent to which surface-density breaks in disc galaxies -- which represent sharp changes in the gradient of the disc density -- drive new spiral modes. We employ linear perturbation theory to investigate how disc breaks alter the eigenmode spectrum of an otherwise pure exponential disc.  We find that the presence of a density break gives rise to a set of new, vigorously growing, modes. For a given multiplicity, these edge modes occur in pairs, with closely separated resonances between each pair.  The growth rate of edge modes decreases when the break is weakened or moved outward to lower-density regions of the disc.  Both down- and up-bending profiles excite edge modes, whose origin can be best understood via the gravitational torques they exert on the underlying disc. When the profile is down-bending (Type II) the faster growing mode is the inner one while in the up-bending (Type III) case the outer mode is faster growing. In both cases the faster growing mode has a corotation almost coincident with the break.  We show that the torques of the edge modes tend to smoothen the break.
\end{abstract}



\section{Introduction}\label{intro}

In early photographic work, \citet{VanDerKruit1979, VanDerKruit1987} found sharp truncations in the exponential profiles of disc galaxies. However, later studies using CCDs revealed that the profiles of disc galaxies are more accurately described by a double exponential, where a break in the disc density is followed by an outer disc region with a smaller scale length \citep{Pohlen+2002}. Subsequently, \citet{Pohlen+2006} and \citet{Erwin+2008} showed that disc profiles come in three types: single exponential down to the last measured point (Type I profiles), downward-bending (truncated or Type II profiles), and upward-bending (anti-truncated or Type III profiles). They also found that, in the local universe, disc galaxies with broken stellar profiles are the norm rather than the exception, with single exponential discs comprising only $\sim 10\%$ of the population. Type II profiles apparently occur in $\sim 40-60\%$ of disc galaxies in the local universe \citep{Pohlen+2006,Erwin+2008}. Typically, in a Type II disc galaxy, the radial scale length drops by a factor of 2 to 3 at the break in the I-band, whereas in a Type III galaxy it increases by a factor of 1.5 to 2 \citep{2016A&A...596A..25L}. Type III breaks appear to occur further out in disc galaxies than Type II breaks, with the typical Type II break radius around 2 inner scale lengths and the typical Type III break radius around 4 inner scale lengths, with substantial scatter \citep{2016A&A...596A..25L}. Type~II breaks have also been observed at high redshift \citep{Perez2004, Xu+2024}, with break radii appearing to have increased with cosmic time \citep{Trujillo+2005, Azzollini+2008b}, linking the evolution of breaks with the inside-out growth of discs. 

Despite their ubiquity, the origin of density-profile breaks is still somewhat uncertain. Type II breaks are generally thought to be due to star formation thresholds \citep{Kennicutt1989, Schaye2004, Elmegreen+2006, Martin+2012}. Numerical simulations of isolated discs support the idea that Type II profiles are the result of a cut-off in the star-formation in combination with radial migration induced by transient spiral arms \citep{Roskar+2008a}. This mechanism predicts a positive stellar age gradient in the disc outside the break, leading to a U-shaped age profile \citep{Roskar+2008a}. Observational studies have also found evidence for U-shaped age profiles \citep{Yoachim+2010,Radburn+2012}. Still other studies find that Type II breaks are not necessarily connected with radial migration, and that disc breaks are largely absent from the mass distribution of the disc, suggesting that the break is caused by differences in stellar populations \citep[e.g.][]{Trujillo+2005, Bakos+2008, Sanchez-Blazquez+2009,Ruiz+2017}. Nonetheless breaks are still observed in resolved star count studies \citep{DeJong+2007, Radburn+2012}
for stellar populations of all ages, albeit weaker in old populations. However, weaker older breaks are required to give the `U'-shaped age profiles.

The cause of Type III (up-bending) profiles is still poorly understood. In some cases, the observed Type III profiles are the result of a transition from the disc to the outer halo \citep{Erwin+2005}, which can be recognized by a change of isophote ellipticity. Studies have proposed various internal \citep[e.g.][]{Minchev+2012, Herpich+2017, Pfeffer+2022} and external mechanisms \citep[e.g.][]{Younger+2007, Kazantzidis+2009, Roediger+2012, Borlaff+2014, Watkins+2019} to account for up-bending profiles. It is also unclear whether the environment plays a role in producing Type~III discs. Cosmological simulations have shown that the incidence of Type~III discs increases with disc mass \citep{Pfeffer+2022}, suggesting mergers are a likely driving mechanism. In contrast, observational studies have found no clear differences in the frequency of break types in cluster and field environments \citep[e.g.][]{Maltby+2012,Head+2015}, which favours internal processes.

While a final theory of the formation of spiral density waves structure is still missing \citep{Lin+1964, Goldreich+1965, Julian+1966, Mark1977, Toomre1981, D'onghia+2013}, both analytic arguments and $N$-body simulations have shown that features such as grooves in the phase space of a disc give rise to a recurrent cycle of spiral modes \citep{Sellwood+1989, Sellwood+1991,DeRijcke+2016}, where a mode is defined as a self-sustaining, sinusoidal disturbance of fixed frequency and constant shape \citep[e.g.][]{Sellwood+2019}. A groove (a narrow deficiency in the phase space density) destabilizes spiral modes, which grow, saturate, and then decay, transferring angular momentum to stars at the outer Lindblad resonance (OLR) in the process. As a result, OLR stars are scattered in a narrow region, carving out new grooves in the disc \citep{Sellwood+2019} and  seeding a next generation of spiral density waves. 
This process repeats itself so long as star formation cools the disc sufficiently to enable it to remain responsive \citep{Sellwood+1984}.

The disc breaks of Type II and III profiles act as discontinuities in phase space and should therefore also produce spiral modes. Already \cite{Toomre1981}, in his analysis of the cold Gaussian disc, found a fast-growing mode (his `D-mode') which he argued was seeded by a sudden drop in the disc density. Consequently, he dubbed this D-mode an `edge mode'. As with groove modes, edge modes do not require any feedback cycles or reflecting barriers to develop. Any infinitesimal co-orbiting distortion at the disc break will induce a response from the surrounding shearing disc in the form of wakes extending inwards and outwards \citep{Goldreich+1965, Julian+1966,  Binney2020}. In the case of Type II profiles, due to the sudden drop in density beyond the edge, the self-gravity of the outer disc offers less support to the outer wake, resulting in an imbalance between the responses of the inner and outer disc. The net forward pull from the inner wake then transports angular momentum outwards beyond the edge, causing the disturbance to grow exponentially as it rotates \citep{Sellwood+2021}.

\citet{Toomre1989} argued that two conditions are necessary for a disc to support edge modes: first, that the disc needs to be massive and cool enough to respond to disturbances in the phase space density, with Toomre-$Q \leq 2$ and $X \equiv \lambda_{\rm \theta}/\lambda_{\rm crit} \leq 3$, where $ \lambda_{\rm \theta} = 2\pi r_{\rm CR}/m$ and the critical wavelength $\lambda_{\rm crit} \equiv 4\pi^{2}G\Sigma/\kappa^{2}$. Here, $r_{\rm CR}$ and $m$ are the disturbance's co-rotation radius and multiplicity, respectively, $\Sigma$ is the disc's surface density and $\kappa$ the epicyclic frequency.
Toomre's second condition is that the radial region over which the disc undergoes its most abrupt change must be smaller than one-quarter of the axisymmetric stability length, $\lambda_{\rm crit}$.

Edge modes driven by Type II and Type III density profiles have received scant attention in the past. Therefore, in this paper, we aim to shed more light on the role of disc breaks (of both Type II and III) in driving edge modes. We use \textsc{pystab} \citep{DeRijcke+2016}, a \textsc{python/C++} code which allows us to calculate the eigenmodes of a razor-thin disc using linear perturbation theory to compute their properties. We present \textsc{pystab} along with the control model we employ in Sec.~\ref{sec:model}. The edge modes driven by the break are presented in Sec.~\ref{sec:indentify_mode}, and the results for a Type~II disc are shown in Sec.~\ref{sec:typeII}. In Sec.~\ref{sec:cool_disc} we explore the effects of varying Toomre-$Q$. Sec.~\ref{sec:mult} discusses differences in the edge mode growth rates. We investigate the properties of edge modes occurring in Type~III discs in Sec.~\ref{sec:type_III}. Lastly, we discuss and summarize our results in Sec.~\ref{sec:disc}.

\section{The models}\label{sec:model}

 \begin{figure*}
	\includegraphics[width=\linewidth]{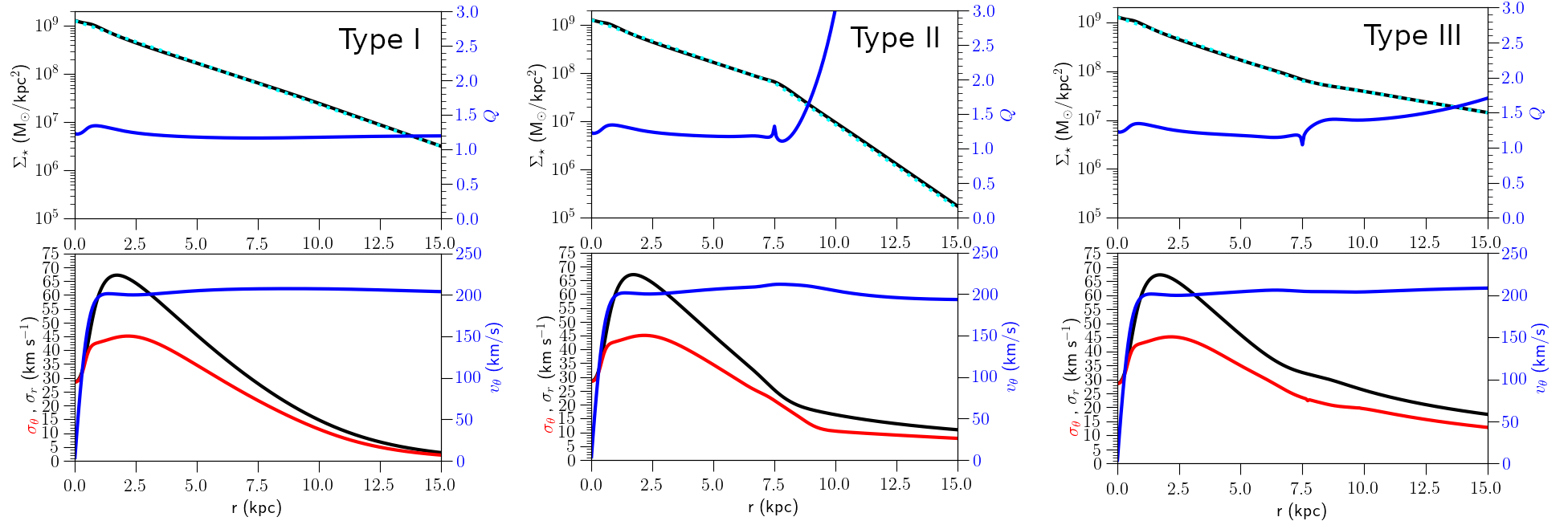}
    \caption{Top panel: The desired surface density distribution (dashed cyan) and the actual density profile produced by the distribution function (solid black) for the control model (left) and a model with a Type~II (middle, with $f_{\rm br} = 0.5$) and Type~III (right, with $f_{\rm br} = 2.0$) break. The blue curve represents the Toomre $Q$-parameter. Bottom panel: the mean tangential velocity $v_{\theta}$ (blue curve), the tangential and radial velocity dispersions, $\sigma_{\mathrm{\theta}}$ (red curve), and $\sigma_{\mathrm{r}}$ (black curve), respectively, for the same models.}
    \label{fig:mom}
\end{figure*}

 \begin{figure}
	\includegraphics[width=\linewidth]{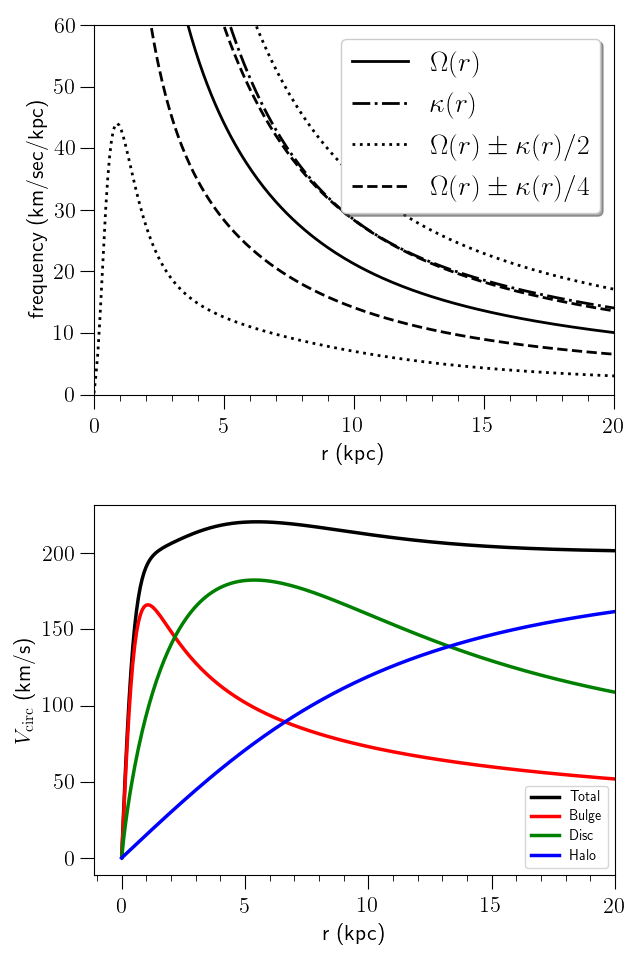}
    \caption{Top: The angular velocity, $\Omega (r)$, epicyclic frequency, $\kappa (r)$ and $m=2$ and $m=4$ Lindblad frequencies, $\Omega (r) \pm \kappa(r)/m$, of the control model. Bottom: The rotation curve of the control model decomposed into the bulge (red), disc (green) and halo (blue). }
    \label{fig:freq}
\end{figure}

 \begin{figure}
	\includegraphics[width=\linewidth]{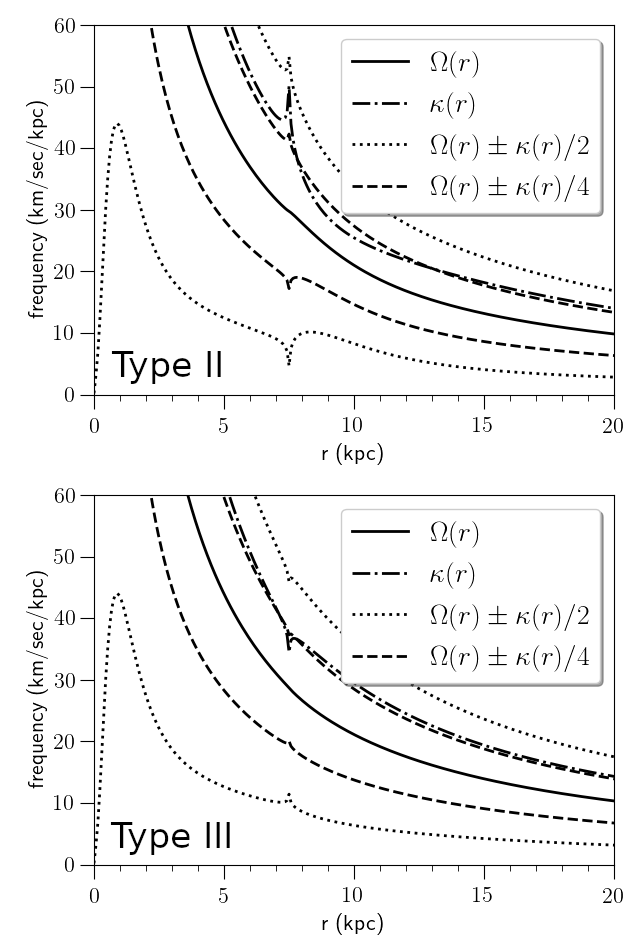}
    \caption{Top: The angular velocity, $\Omega (r)$, epicyclic frequency, $\kappa (r)$ and $m=2$ and $m=4$ Lindblad frequencies, $\Omega (r) \pm \kappa(r)/m$, of the Type II (top) and Type III (bottom) model. The bumps in the frequency profiles at the break radius are a direct result of the abrupt change in the derivative of the density at that radius.}
    \label{fig:freqII}
\end{figure}

The control model comprises a razor-thin, pure exponential stellar disc with a density distribution given by:
\begin{equation}\label{eqn:disc}
   \centering
   \Sigma = \Sigma_{0} e^{-r/h_\mathrm{d}},
\end{equation}
where $\Sigma_{0}$ is the central stellar surface density and $h_{\rm d}$ is the scale length of the disc. We take $h_{\rm d}=2.5$~kpc \citep[as appropriate for the Milky Way, see][]{Bland-Hawthorn+2016} and adjust $\Sigma_0$ such that the total disc mass is $M_{\rm d}=5\times 10^{10}~M_\odot$. We compute the in-plane gravitational potential, $V_{\rm{d}}(r)$, of the stellar disc numerically; we use the same code to compute the potential of the disc when a density break is introduced. We employ a Plummer sphere for the bulge and a logarithmic potential for the halo. The in-plane gravitational potentials of these non-responsive components are given by Equations \ref{eqn:bulge} and \ref{eqn:halo}, respectively:
\begin{align}
    V_{\mathrm{b}}(r) &= -\frac{GM_{\rm b}}{\sqrt{r^2+r_{\rm b}^2}}, \label{eqn:bulge} \\
    V_{\mathrm{h}}(r) &= \frac{v^2_0}{2}\ln \left( r_{\rm h}^2+r^2 \right). \label{eqn:halo}
\end{align}
The bulge has a mass $M_{\rm b}=M_{\rm d}/4$ and scale-length $r_{\rm b}=0.3 h_{\rm d}$. The halo scale length is $r_{\rm h}=5r_{\rm d}$ and $v_0 = 0.65 \sqrt{GM_{\rm d}/h_{\rm d}}\approx 191$~km~s$^{-1}$.

The equilibrium configuration of the model is characterized completely by the global potential $V_0(r)=V_{\rm{d}}(r)+V_{\mathrm{b}}(r)+V_{\mathrm{h}}(r)$ and the distribution function (DF) of the disc. We use the DF derived by \citet{Dehnen1999} for a warm disc:
\begin{equation}
   \centering
   f_{0}(E,L)= \frac{\gamma(r_{E})\Sigma(r_{E})}{2\pi\sigma^{2}_{r}(r_{E})} \exp \bigg[  \frac{\Omega(r_{E})[L-L_{c}(E)]}{\sigma^{2}_{r}(r_{E})} \bigg],
\end{equation}
where $r_E$ is the radius of the circular orbit at energy $E$, and $\gamma \equiv 2\Omega/\kappa$, with $\Omega$ the circular orbit angular frequency. $L$ and $E$ are the angular momentum and binding energy, respectively, and are given by:
\begin{align}
    L &= rv_{\theta}, \\
   E &= V_{0}(r)-\frac{1}{2}(v^2_r+v^2_{\theta}),
\end{align}
$v_r$ and $v_\theta$ being the radial and tangential velocity, respectively. Using a DF of this form, we can generate disc galaxy models with any desired surface density profile, $\Sigma(r)$, and radial velocity dispersion profile, $\sigma_r(r)$. This allows us to introduce breaks into the surface density of the control model; the cooler the orbital distribution, the closer a model adheres to the target density and dispersion profiles. Small deviations between the model's characteristics and the target profiles can occur if the target profiles have extremely sharp features. This limits the sharpness of the density breaks we can introduce, but has no further consequences for our analysis. 

In the top left panel of Fig.~\ref{fig:mom}, we show the disc's density (black) and Toomre $Q$ parameter (blue) profiles for the control model with no break (Type I). The dashed cyan line represents the desired profile, while the solid black line reflects the actual density profile fit by the DF. In the control model, $Q \approx 1.2$ and is roughly constant throughout the disc. In the bottom left panel of Fig.~\ref{fig:mom}, we show the radial and tangential velocity dispersions, as well as the mean tangential velocity curve of the control model\footnote{These profiles are computed via numerical integrals of the DF.}. Note that the velocity dispersion profiles only reflect the contribution from the disc, causing a central dip at $r<2\kpc$. The bulge in the model -- which is only included via its potential -- will result in a $\sigma_r$ profile which decreases with $r$, consistent with observations \citep[e.g.][]{Mogotsi+2019}.

In the top panel of Fig.~\ref{fig:freq}, we show the angular velocity, $\Omega (r)$, epicyclic frequency, $\kappa (r)$ and $m=2$ and $m=4$ Lindblad frequencies, $\Omega (r) \pm \kappa(r)/m$, of the control model. The $\Omega (r) - \kappa(r)/2$ curve, which determines the location of the inner Lindblad resonance, or ILR, peaks at roughly $45\,\rm{km} \,\rm{s}^{-1} \rm{kpc}^{-1}$. The bottom panel shows the rotation curve of the control model decomposed into its bulge (red), disc (green) and halo (blue) components.

We introduce density breaks in the control disc at radius $r_{\rm br}$ by changing the target density profile, $\Sigma(r)$, in the DF. The target density declines exponentially with radius with a scale length $h_{\rm d,inner}$ inside $r_{\rm br}$ and with a scale length $h_{\rm d,outer}=f_{\rm br} h_{\rm d,inner}$ outside $r_{\rm br}$. We refer to $f_{\rm br}$ as the break strength, since $f_{\rm br} = 1$ corresponds to no break. A Type~II (down-bending) break corresponds to a break strength $f_{\rm br}<1$ while a Type~III (up-bending) break is introduced by adopting $f_{\rm br}>1$. In Fig.~\ref{fig:mom}, we show the properties for a model with a Type~II break with $f_{\rm br}=0.5$ (middle), and a Type~III break with \fbr\ $= 2.0$ (right). Both models have $\rbr=7.5\kpc$. A density break, through its effect on the circular velocity profile, introduces a small kink in the $Q$-parameter, which also causes the velocity dispersion to be slightly higher than the control model around \rbr. In the Type~II model (middle), the drop in density also results in a sharp increase in the $Q$-parameter at $r>\rbr$. The $Q$-parameter also rises steadily at $r>\rbr$ for the Type~III model. This is due to an increase in the stellar epicyclic frequencies at larger radii, which mimics a more heated outer disc. Fig.~\ref{fig:freqII} also shows the epicyclic and Lindblad frequencies for the Type II (top) and Type III models. While the frequencies are qualitatively the same as the control model, the abrupt change in density causes a kink to appear at the break radius. This is a direct result of the abrupt change in the derivative of the density at that radius. Note that for Type~III profiles, the density profile is also always monotonically decreasing with radius. In the rest of this paper, we always set $h_{\rm d,inner}=2.5\kpc$, \ie\ the same as the control model.

We compute the eigenmodes of a given disc model using \textsc{py\-stab}, a fast and versatile \textsc{python/C++} code.  The underlying mathematical formalism of this code is described in \citet{Vauterin+1996}, \citet{Dury+2008}, and \citet{DeRijcke+2016}, so we do not include a detailed description here. Instead, a brief outline of the formalism, including the values and numerical parameters we employ in the models, is given in  Appendix~\ref{pystab}. The bulge and halo components defined above are considered to be too dynamically hot to support any instabilities and are included only via their potentials. Any eigenmodes which \textsc{pystab} recovers exist entirely within the razor-thin disc. This version of the code does not take into account any dynamical effects due to gas.

\begin{figure*}
	\includegraphics[width=1.0\linewidth]{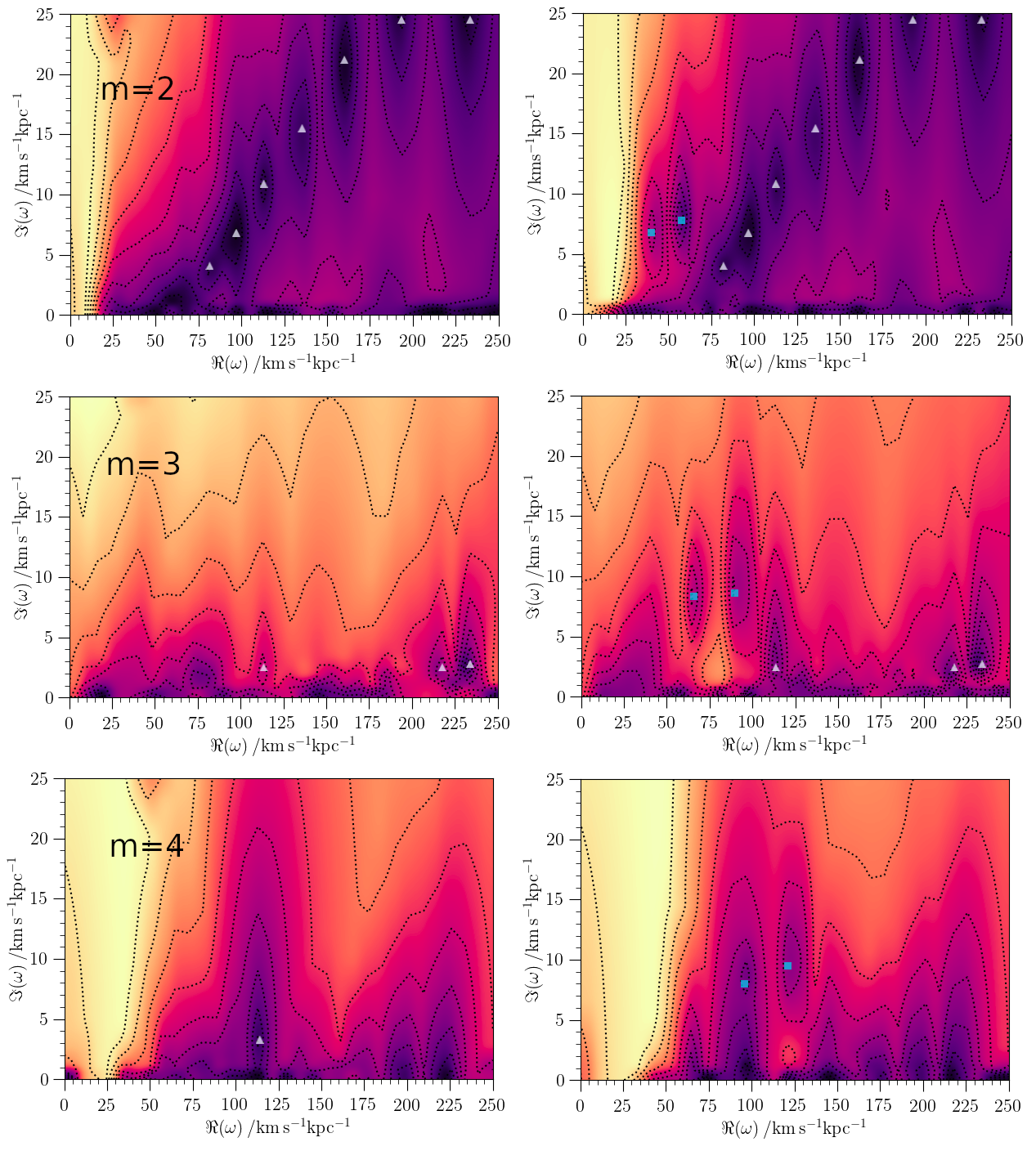}
    \caption{The eigenmode spectra in the complex frequency space of the control model (left column) and the Type~II model with \rbr\ $= 7.5\kpc$ and \fbr\ $= 0.5$ (right column). Each row represents spectra of different multiplicities, with top, middle, and bottom rows showing the spectra for $m=2, ~3$, and $4$ respectively. The real part of the frequency is related to the pattern speed of the mode by $\Omega_{\rm p} = \Re(\omega)/m$, and the imaginary part quantifies the growth rate of the mode. White triangles mark the modes present in the control model, whereas the cyan squares reflect the position of the edge modes. Comparing the two columns for each row shows that a disc break gives rise to a set of edge modes which are absent in the control model.}
    \label{fig:control}
\end{figure*}

\section{The spectrum of the control model}\label{sec:indentify_mode}

In order to identify the edge modes, we first perform the stability analysis of the control model described in Sec.~\ref{sec:model}. This provides us with a basis for comparison when we later analyse the eigenmode spectra of the models after we introduce a disc break. The left-hand column of Fig.~\ref{fig:control} shows the spectra of the control model for different multiplicities. The vertical axis represents the imaginary part of the complex frequency, $\Im({\omega})$, \ie\ the growth rate of the eigenmode. On the horizontal axis, we plot the real part of the complex frequency, which is given by $\Re(\omega) = m\Omega_{\mathrm{p}}$, where $m$ is the multiplicity and $\Omega_{\mathrm{p}}$ is the pattern speed. The different rows show the spectra for $m = 2,~3$, and $4$. 

As explained in Appendix~\ref{pystab}, an eigenmode occurs at each complex frequency $\omega$ where the model-dependent response matrix $C(\omega)$ has a unit eigenvalue. The plots of the spectra therefore show the value of $\min(|\lambda-1|)$: the smallest distance between 1 and any of the eigenvalues of $C(\omega)$. In Fig.~\ref{fig:control} and its analogues, eigenmodes can be identified visually as the dark regions where $\min(|\lambda-1|)=0$. The eigenmodes which are present in the control model are marked with white triangles.

The top left panel of Fig.~\ref{fig:control} shows that the control model supports a sequence of growing $m = 2$ eigenmodes. These eigenmodes have pattern speeds $\Omega_{\rm p} \gtrsim 50\,\rm{km} \,\rm{s}^{-1} \rm{kpc}^{-1}$ and rotate sufficiently fast to avoid having an ILR (see Fig.~\ref{fig:freq}). Therefore, these are likely to be cavity modes which reflect between the galaxy centre and another resonance. In order to verify this, we set up a model with an increased bulge mass ($M_b = M_d/1$), which effectively raises the peak of the ILR in the model. We find that the sequence of  $m=2$ modes is suppressed, which is expected since cavity modes are damped by the ILR \citep{Mark1974}.

 The complex frequencies and resonances of all the $m = 2$ eigenmodes in the control model are listed in Table~\ref{table:1}. In the $m=3$ case of Fig.~\ref{fig:control}, we find three slowly growing cavity modes with varying pattern speeds. The third row of Fig.~\ref{fig:control} shows that the control model supports a single slow-growing $m=4$ eigenmode at $m\Omega_{\rm p}\approx 112\,\rm{km} \,\rm{s}^{-1} \rm{kpc}^{-1}$. 

\begin{figure*}
	\includegraphics[width=1\linewidth]{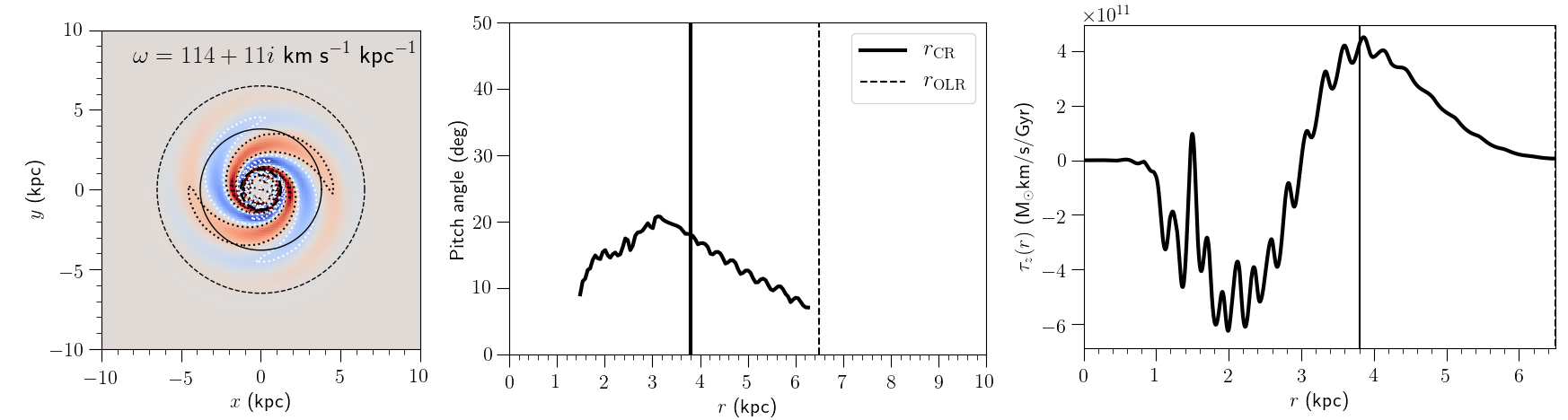}
    \caption{Left: the density distribution of an intrinsic mode of the disc, where the solid curve represents the CR radius and the dashed curves represent the ILR and OLR radii. Middle: the pitch angle of the mode. Right: The torque per unit width $dr$ exerted by the mode as a function of radius.}
\label{fig:control_mode}
\end{figure*}

In Fig.~\ref{fig:control_mode}, we show some characteristics of one of the intrinsic eigenmodes of the control model. We chose the $m=2$ mode with complex frequency $\omega=114+11i\,\rm{km} \,\rm{s}^{-1} \rm{kpc}^{-1}$ as representative of the modes in the control model. The left panel shows the surface density of the mode, where the solid circle represents the corotation (CR) radius, while the dashed circles represent the inner and outer Lindblad resonances. Positive densities are indicated in red, while negative ones are shown in light blue. The middle panel shows the pitch angle of the spiral as a function of radius. (The wiggles in the pitch angle profiles are numerical artefacts due to the use of a finite set of potential-density basis pairs, as described in Appendix~\ref{pystab}).

Lastly, the right panel of  Fig.~\ref{fig:control_mode} shows the gravitational torque, $\tau_z(r)$, exerted by the spiral pattern on an annulus with radius $r$ and unity width. The torque on an annulus of width d$r$ is given by
\begin{equation}\label{eq:torque}
    \tau_z(r) = m  \Sigma_{\text{pert}}(r) V_{\text{pert}}(r) \pi r dr
    \sin(m\gamma_0(r)), 
\end{equation}
where $\Sigma_{\text{pert}}$ and $V_{\text{pert}}$ are the amplitudes of the pattern's density and potential spirals, respectively, $\gamma_0$ is the angular offset between those two spirals, and $m$ is the pattern's multiplicity \citep[see][]{Zhang+1996, Zhang+1998}. As a sanity check, we confirmed that the total torque on the disc, $\tau_{z,\mathrm{total}}$, is zero, and this for all modes discussed in this paper. In other words, we numerically checked that the integral 
\begin{equation}
    \tau_{z,\mathrm{total}} =  \int_0^\infty m \pi r \Sigma_{\text{pert}}(r) V_{\text{pert}}(r)
    \sin(m\gamma_0(r)) dr
\end{equation}
is indeed zero, as it should.

A negative $\tau_z(r)$ means that the disc stars lose angular momentum to the spiral mode, whereas a positive $\tau_z(r)$ indicates that they gain angular momentum from the pattern. As expected, the outer stars gain angular momentum at the expense of the inner ones. This outward transport of angular momentum is the root cause of spontaneously growing eigenmodes \citep[see][]{1972MNRAS.157....1L,1998ApJ...499...93Z}.

\begin{figure}
	\includegraphics[width=1\linewidth]{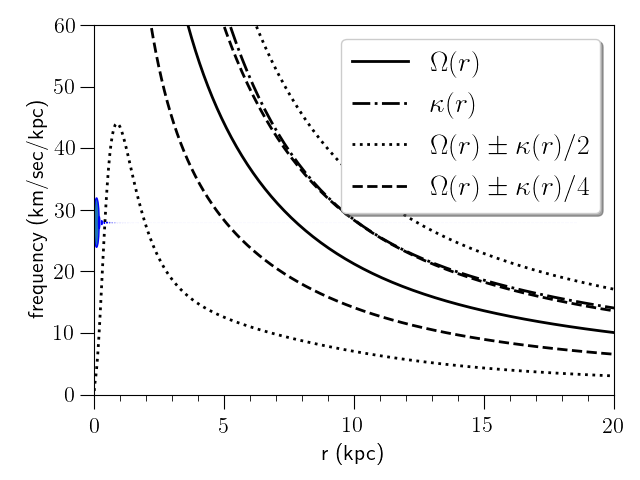}
    \caption{The important frequencies of the control model (same as Fig.~\ref{fig:freq}). The blue region at small $r$ is centred around the frequency of a highly centrally concentrated $m=4$ mode with $m\Omega_{\rm p}\approx 112\,\rm{km} \,\rm{s}^{-1} \rm{kpc}^{-1}$ found in the control model. The vertical width of the region -- which is proportional to its density -- reflects the amplitude of the wave. This is not connected to the values on the y-axis, and is only indicative. We find a number of these highly concentrated modes across our study. However, since they do not impact the evolution of the outer disc, we omit them from further study.}
\label{fig:ilr_mode} 
\end{figure}

\subsection{Highly concentrated modes}\label{sec:conc_mode}

In addition to the cavity modes described above, we uncover a number of highly radially concentrated modes living entirely within their ILR. In the $m=4$ case (bottom row of Fig.~\ref{fig:control}), we find such a mode around $m\Omega_{\rm p}\approx 112\,\rm{km} \,\rm{s}^{-1} \rm{kpc}^{-1}$. In Fig.~\ref{fig:ilr_mode}, we show the radial extent of this mode. The blue region in the plot is centred about the frequency of the mode and its vertical width is proportional to its density, which reflects the amplitude of the wave. The width of the blue region is not linked with values on the y-axis and is only indicative. As can be seen, the mode's amplitude is entirely restricted to a region within $500\pc$ of the centre, which is inside the ILR. This type of mode may be similar in nature to that identified by \citet{Binney2020} in their study of the shearing sheet. However, given how centrally concentrated these modes are -- existing in a region smaller than the thickness of an average disc -- they are unlikely to play any significant role in the overall evolution of the outer disc. 

In addition, we find that in some of the models we explore below, the density break also destabilizes one of the concentrated modes described. However, it is unaffected by any changes to the break properties. That is, its pattern speed and growth rate remain fixed. Therefore, we shall omit these modes from further analysis. 

We found that, in some cases, there are other modes in the complex frequency being masked by this concentrated mode. Therefore, whenever we find a concentrated mode, we carefully verify that no extended modes are masked by this concentrated mode.

\begin{figure*}
	\includegraphics[width=1.0\linewidth]{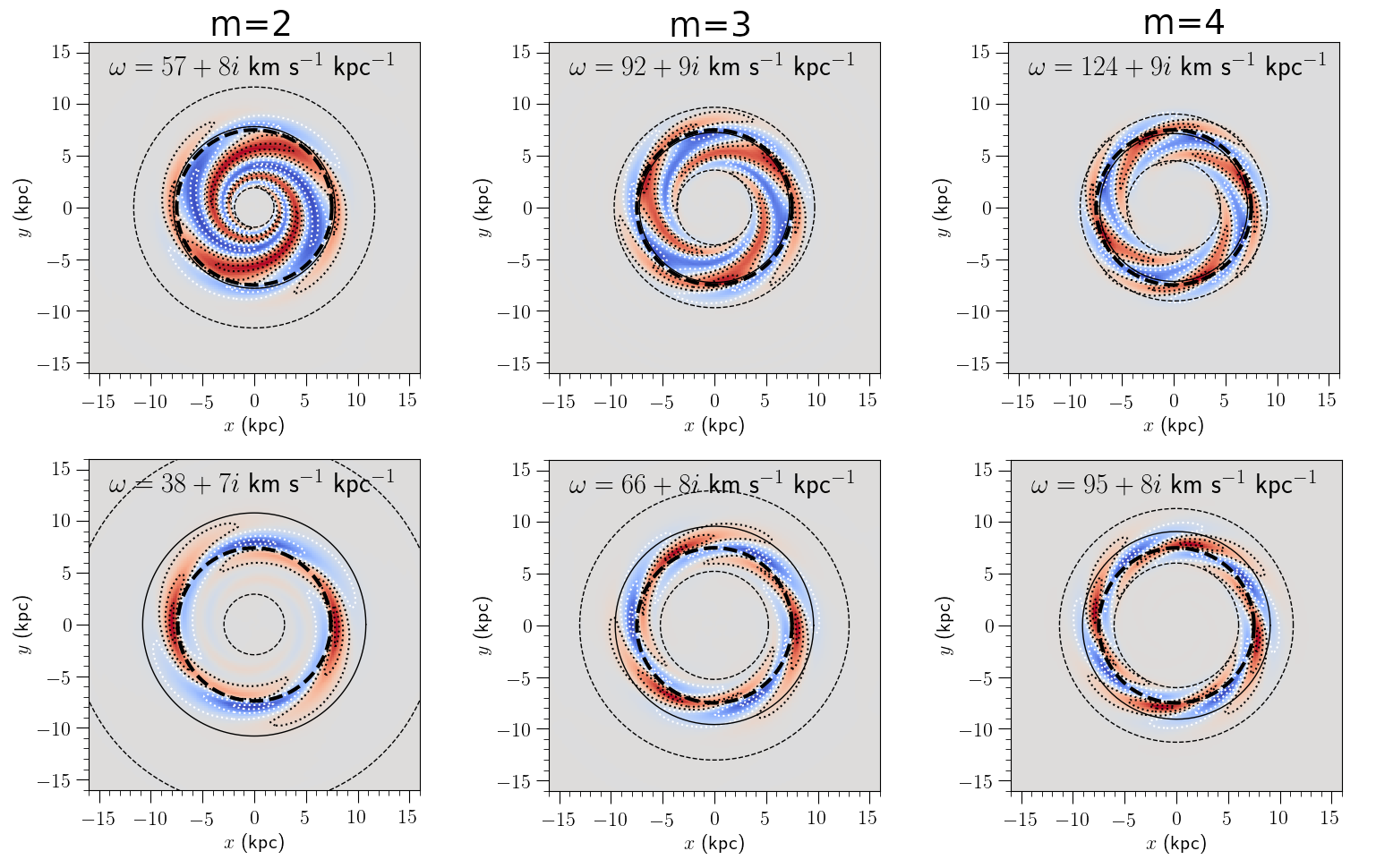}
    \caption{Surface density of the edge modes for the model with \fbr\ $=0.5$ and \rbr\ $= 7.5 \kpc$, with each column representing modes of different multiplicities, $m$. Positive densities are shown in red, negative ones in blue. The disc break radius is indicated by the thick dashed circle, while the thin dashed circles indicate the positions of the inner and outer Lindblad resonances of circular orbits. The solid line represents the corotation resonance of the mode. Note the near coincidence of the OLR of the fast mode (top row) with the CR of the slow mode (bottom row). The thin dotted line contours trace surface density levels at $\pm 10$, $\pm 50$, and $\pm 90$ per cent of the maximum value}
    \label{fig:denmode}
    
\end{figure*}

\begin{table}
\centering
    \begin{tabular}{ cccc }
        \multicolumn{4}{c}{$m = 2$ control disc model  modes} \\
        \hline
        $\Re({\omega})$ (km/s/kpc)  & $\Im({\omega})$ (km/s/kpc) & CR (kpc) & OLR (kpc)\\
        \hline
        233.871 & 24.366 &1.748 & 3.189\\
        193.518 & 24.544 & 2.156 & 3.914\\
        161.290 & 21.452 & 2.645 & 4.733\\
        137.733 & 14.676 & 3.210 & 5.625\\
        114.903 & 10.562 & 3.801 & 6.508\\
        96.771 & 6.77 & 4.893 & 8.073\\
        81.531 & 3.694 & 5.41 & 8.81\\
        \hline
    \end{tabular}
\caption{Complex frequencies, and CR and OLR radii of the $m = 2$ modes of the control model (a single exponential profile). The modes are listed in order of decreasing growth rate.}
\label{table:1}
\end{table}

\begin{table}
\centering
    \begin{tabular}{ ccccc }
        \multicolumn{5}{c}{Edge mode frequencies for $r_{\mathrm{br}} = 7.5\kpc$ \& $f_{\mathrm{br}} = 0.5$} \\
        \hline
        $m$ & $\Re({\omega})$ (km/s/kpc) & $\Im({\omega})$ (km/s/kpc) & CR (kpc)& OLR (kpc)\\
        \hline
        2 &  57.177 & 7.903 & 7.821 & {\bf 11.650} \\
        2 &  37.709 & 6.756 & {\bf 10.909} & 17.827 \\
        3 &  91.814 & 8.734 & 7.257 & {\bf 9.691}\\
        3 &  66.478 & 8.202 & {\bf 9.591} & 13.023 \\
        4 & 124.108 & 9.390 & 7.150 &  {\bf 9.040}\\
        4 &  95.147 & 8.419 & {\bf 9.076} & 11.302\\
        \hline
    \end{tabular}
\caption{Complex frequencies, and CR and OLR radii of the modes of a model with a density break with $r_{\mathrm{br}}~=~7.5\kpc$ and $f_{\mathrm{br}}~=~0.5$. The first column lists the multiplicity $m$ of the modes. The CR and OLR of the two members of each mode pair which show hints of resonance overlap are highlighted in bold-face.}
\label{table:2}
\end{table}

\section{Edge modes in Type II discs}\label{sec:typeII}

We first consider the edge modes in Type~II discs. The right column of Fig.~\ref{fig:control} shows the eigenmode spectra for a model with a Type~II disc break with $r_{\rm br} = 7.5\kpc$ and $f_{\mathrm{br}} = 0.5$. Comparing the left and right columns of Fig.~\ref{fig:control}, it is clear that the appearance of the disc break alters the disc's eigenmode spectrum and gives rise to a new set of rapidly growing modes, not present in the control model, with different pattern speeds for all multiplicities (\ie\ for all rows). These new modes are marked with cyan squares in Fig.~\ref{fig:control}. Unlike the $m=2$ cavity modes in the control model, we find that the growth rate of the edge modes are not affected by an increase in bulge mass, despite having an ILR. For all $m$ (all rows), edge modes occur in pairs, with both members of each pair having comparable growth rates (see Table.~\ref{table:2}), with the slow (outer) mode having a growth rate on average $10\%$ lower than the fast (inner) member. We list the complex frequencies of these edge modes, along with additional parameters, in Table~\ref{table:2}. The density break clearly has a substantial destabilizing effect on the eigenmode spectrum, altering the number and the frequencies of the eigenmodes.

In this particular model, with $r_{\mathrm{br}} = 7.5\kpc$ and $f_{\mathrm{br}} = 0.5$, we find that the $m = 4$ mode is the fastest growing one (see Table \ref{table:2}) and would therefore dominate the evolution of the galaxy. While we only show our results for modes up to $m = 4$, we have also tested the model for higher multiplicity edge modes and found similar growth rates up till $m = 8$. Beyond $m = 8$, the growth rates steadily decline, with the $m = 16$ edge modes having growth rates under $5 \kms\kpc^{-1}$. Since we are working within the linear framework, higher multiplicity modes are not the result of overlapping lower multiplicity modes. Throughout the study, we find that the characteristics we will describe are similar for all edge modes, regardless of the multiplicity. Therefore, for simplicity, we restrict the rest of our analysis to the leading modes with $m \leq 4$.

In Fig.~\ref{fig:denmode}, we plot the surface density of the edge modes, with their complex frequency denoted in each panel. Here, the solid black curve represents the corotation radius, while the thick dashed curve reflects the break radius. The inner and outer thin dashed lines represent the inner and our Lindblad resonances, respectively. As can be seen in Fig.~\ref{fig:denmode}, the inner fast member (top row) of each mode pair has its CR roughly coincident with the break. In addition, it is clear that the wave is unable to propagate through the ILR. On the other hand, the outer slow member (bottom row) has its CR well outside the break. For all $m$, the outer edge mode exists between the break and its CR.

In addition, we find that for all $m$, the outer Lindblad resonance (OLR) of the fast mode almost coincides with the CR of the slow mode (see Table \ref{table:2}). This suggests the two modes are resonantly coupled, which may drive more complex non-linear behaviour \citep[e.g.][]{Sygnet1988, Masset+1997}. We explore this in more detail in Sec.~\ref{sec:coupling}. Parallels may be made between our results and the characteristics of the groove modes reported in \citet{DeRijcke+2016}; their linear stability analysis of a grooved exponential disc revealed that spiral modes triggered by a groove also occur in pairs, and that there is the possibility of resonant overlap. This is unsurprising since grooves and edges are closely related \citep[e.g.][]{Sellwood+2021}.

\begin{figure*}
	\includegraphics[width=1\linewidth]{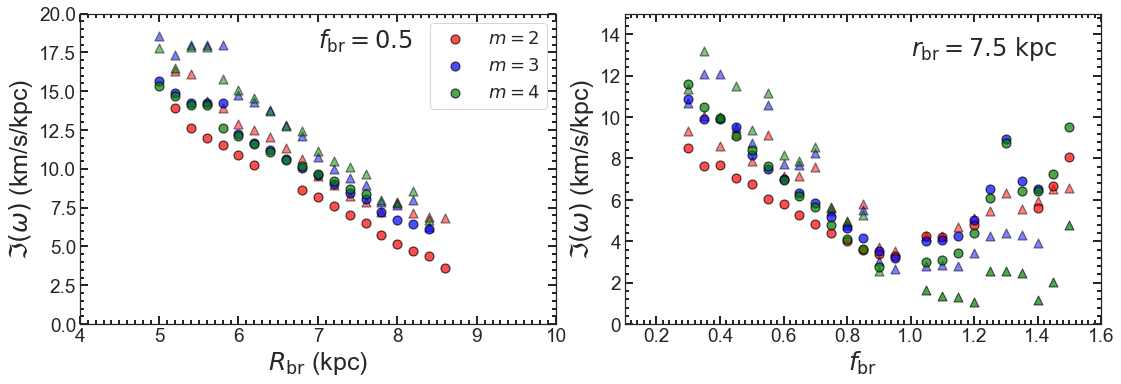}
    \caption{Left: The growth rate, $\Im({\omega})$, of the edge modes as a function of the break radius, \rbr, for a model with \fbr\ $= 0.5$ (Type~II profile). The $m=2$, $3$, and $4$ modes are represented by red, blue, and green markers respectively, with each separate pair of modes represented by circle (slow, outer mode) and triangle (fast, inner mode) markers. For all values of $m$, the growth rate of the edge modes decreases as \rbr\ increases, until the edge modes become lost and only the intrinsic modes of the disc remain. Right: The growth rate of the edge modes as a function of the break strength, \fbr, for models with $r_{\rm br}= 7.5 \kpc$. Here, \fbr\ $ = 1$ represents the control model (no break), while $f_{\rm br} < 1$ and $f_{\rm br} > 1$ represent Type~II and Type~III breaks, respectively. As $f_{
     \rm br}\rightarrow 1$ from smaller values, the Type~II break weakens, which causes the edge modes to grow less rapidly. As $f_{\rm br}$ increases past unity, the models transition to Type~III profiles, which revives the edge modes.}
\label{fig:rbr} 
\end{figure*}

\subsection{Dependence on break radius}\label{sec:break_radius}

We investigate the impact of varying the break radius, \rbr, on the growth rates of edge modes in a Type II disc, by varying it in the range $5 \leq \rbr/\kpc \leq 9$ in steps of $0.2\kpc$, holding $f_{\rm br} = 0.5$ fixed. We compute the eigenmode spectrum in each instance; the resulting growth rates, $\Im({\omega})$, are shown in the left panel of Fig.~\ref{fig:rbr}. The modes with multiplicities $m=2,~3$, and $4$ are shown in red, blue and green markers, respectively, with the two members of each mode pair represented by circles for the outer (slow) mode and triangles for the inner (fast) mode. For all values of \rbr, the $m=2$ mode is the slowest growing mode.

Fig.~\ref{fig:rbr} (left) reveals that all multiplicities exhibit the same trend: as \rbr\ increases, the growth rates for the edge modes decline. At $\rbr \gtrsim 9\kpc$ only the intrinsic modes of the control disc (left panel of Fig.~\ref{fig:control}) remain visible.
The dependence of the growth rate on {\rbr} is easy to understand in the edge mode mechanism of \citet{Toomre1989}, in which the growth of the edge modes depends entirely on the supporting response of the disc. The less support the surrounding disc provides to the wakes generated by an initial perturbation on the axisymmetric break, the weaker are the resulting torques on the disc, which in turn lowers the growth rate of the modes.
The rate at which the growth rate of the edge modes declines is roughly equal for all values of $m$.

\subsection{Dependence on break strength}

\begin{figure*}
	\includegraphics[width=1.0\linewidth]{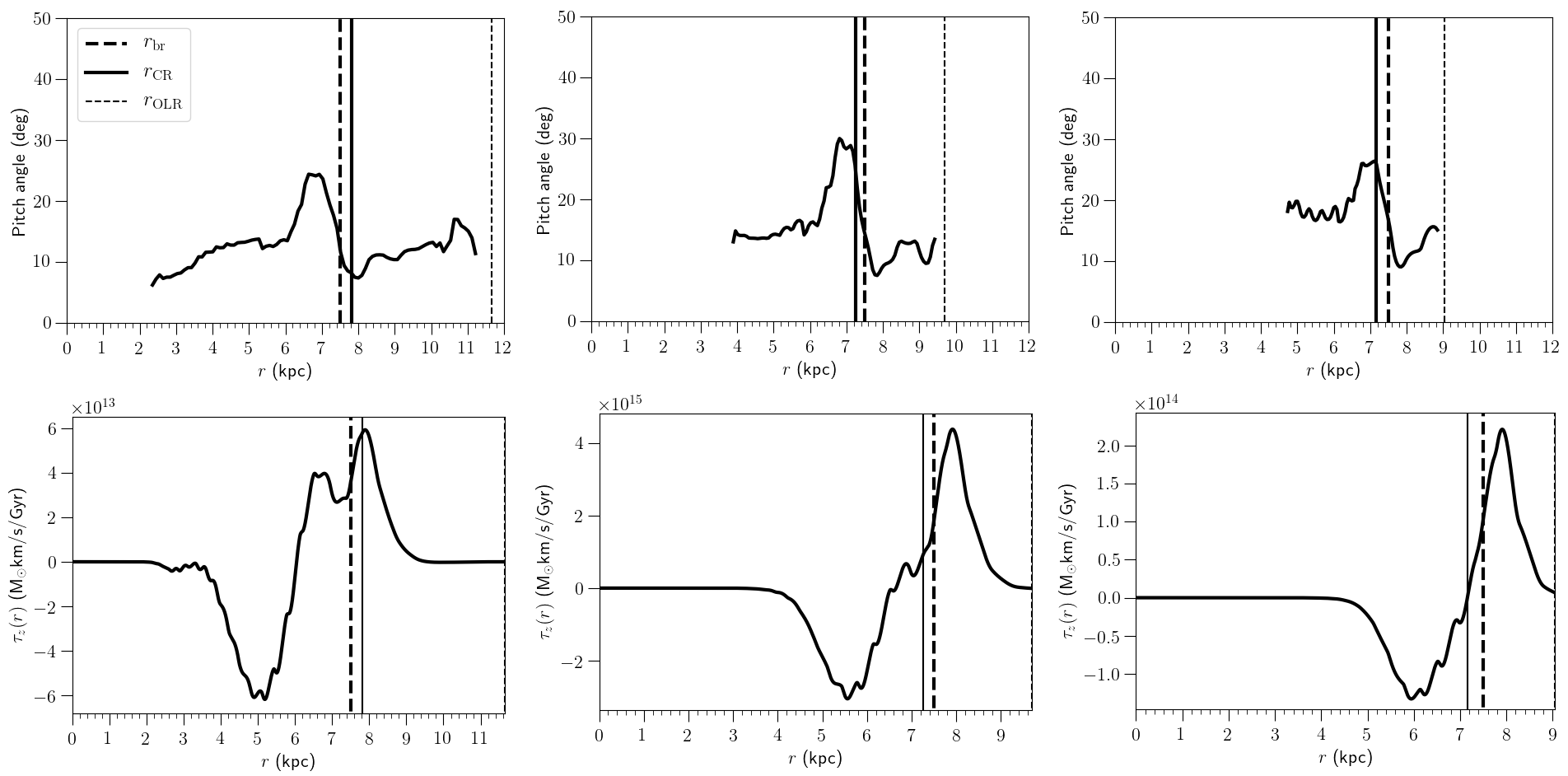}
    \caption{The pitch angles (top row) and torques (bottom row) of the edge modes for the model with \fbr\ $=0.5$ and \rbr\ $= 7.5 \kpc$. We show only the fast modes, which have their CR close to the disc break (top row in Fig.~\ref{fig:denmode}). The left, middle, and right columns show the results for $m=2$, $3$, and $4$, respectively. The vertical full line marks the location of each mode's CR radius; the vertical dashed line marks the break radius.
    }
    \label{fig:pitch}
\end{figure*}

\begin{figure}
	\includegraphics[width=1.0\linewidth]{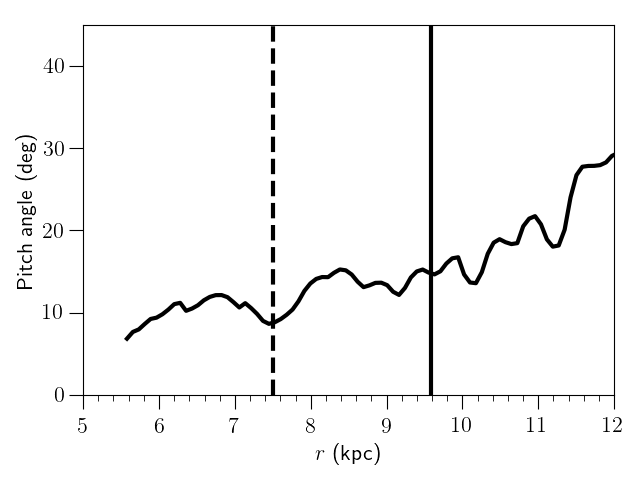}
    \caption{The pitch angle as a function of radius $r$ for the outer (slow) $m=3$ edge mode for the model with \fbr\ $=0.5$ and \rbr\ $= 7.5 \kpc$. For all $m$, the pitch angle variation of the outer edge modes show similar behaviour, with the pitch angle increasing steadily with radius.
    }
    \label{fig:outer_pitch}
\end{figure}

We next investigate the impact which the break strength, \fbr, has on the growth rates of the edge modes, by varying the scale length of the outer disc, while holding $\rbr\ = 7.5 \kpc$ fixed. The case of $f_{\rm br}=1$ represents the control model with no break (equivalent to the left column of Fig.~\ref{fig:control}), while $f_{\rm br}<1$ and $f_{\rm br}>1$ correspond to Type~II and Type~III profiles, respectively. 

The right panel of Fig.~\ref{fig:rbr} plots the edge mode growth rate, $\Im(\omega)$, as a function of \fbr. The break strength has a dramatic impact on the growth rate of the edge modes. As with varying \rbr, the same trend is present for all values of $m$: as the break weakens ($f_{\rm br} \rightarrow 1$) a substantial decrease in the growth rate of the edge modes occurs, until only the intrinsic modes of the control model remain. As \fbr\ increases past unity, the model transitions from a Type~II to a Type~III break, with a revival of the edge modes; we defer discussion of Type~III breaks to Section~\ref{sec:type_III}.

The gradual weakening of the edge modes as the Type II break is made progressively gentler can be attributed to both the weakening of the density contrast across the break and to the increase in the outer disc mass. In the context of the Toomre mechanism, as the density contrast across the break radius becomes smaller, and finally disappears when $f_{\rm br}=1$, the break becomes an ever weaker source of initial perturbations -- stunting the growth of the edge modes. In addition, as the outer disc gains mass, it supports a stronger outer wake from an initial perturbation, and the relative difference in the responses of the inner and outer discs diminishes. As a result, the growth rates of the edge modes decrease until $f_{\rm br}=1$ is reached and only the modes intrinsic to the disc remain.

We conclude that both the strength of the break and its radial position are important factors which contribute to the growth rates of the edge modes. Even a weak break produces edge modes as long as the break radius lies in a responsive part of the disc. However, for a weak break, the edge modes will likely not be dynamically important, and will be overwhelmed by other more dominant modes of the disc. On the other hand, a strong break leads to vigorously growing edge modes (especially for $m = 3\, \text{and} \, 4$), which may play a role in the evolution of the galaxy.

\subsection{Pitch angles and torque}\label{sec:torque&pitch}

Fig.~\ref{fig:control_mode} shows the pitch angle (middle panel) and torque (right panel) profiles for one of the $m=2$ cavity modes in the control model. The pitch angle increases up to $\sim 3\kpc$, followed by a steady decrease beyond. We find that this smooth variation in the pitch angle is similar for all cavity modes.

However, in the case of edge modes, we find that the pitch angle variation is unlike that of the cavity modes. In Fig.~\ref{fig:pitch}, we show the pitch angle (top row) and the torque on the stars, $\tau_z(r)$, (bottom row) of the  fast edge modes, as a function of radius for $m = 2,\, 3,\, \text{and}\, 4$ (left to right). We do this for a model with \fbr\ $=0.5$ and \rbr\ $=7.5 \kpc$. Comparing the pitch angle profiles here with that of the control model in Fig.~\ref{fig:control_mode} shows that while the pitch angle for the control model mode varies smoothly over a wide radial range, the change in the edge mode pitch angles is more abrupt. For example, the pitch angle for the $m=2$ (left panel of In Fig.~\ref{fig:pitch}) does not show significant variation away from the break region. However, right inside the break, the pitch angle increases rapidly (by $\approx 10\degrees$), then falls back to roughly its original values immediately outside the break region, resulting in a tightening of the spiral pattern beyond \rbr, as can also be seen in the top row of  Fig.~\ref{fig:denmode}. The pitch angles of the $m=3$ and $m=4$ modes behave similarly. 

The sudden change in pitch angle for the edge modes also has an effect on the torque the spiral produces, with more tightly wound edge modes exerting their torque over narrower radial ranges. We show the corresponding torque profiles in the bottom row of Fig.~\ref{fig:pitch}. Generally, the torque profiles follow similar overall behaviour for $m=2,3$, and $4$. Inside the break (and CR), the torque is negative, whereas it is positive outside the break. Typically, the region around a Type~II break lies on the rising side of the positive part of the $\tau_z(r)$ profile. In other words, this part of the disc gains angular momentum and is therefore being stretched outwards. Moreover, the stars just inside the break gain less angular momentum than the stars just outside the break. Hence, stars are being pulled away from the break, spreading the mass outwards and decreasing the sharpness of the break. Conversely, stars inside the break (and CR) lose angular momentum due to the negative torque and are therefore pushed inwards. This results in the inner density profile becoming steeper over time, weakening the break further. The combined effect of this secular evolution driven by edge modes is that the break will become weaker and thus less effective at provoking edge modes.

Fig.~\ref{fig:outer_pitch} shows the pitch angle variation for one of the slow, outer edge modes ($m=3$). We find similar behaviour for all values of $m$. The pitch angle for the slow edge modes differs substantially from its fast counterpart. Rather than undergoing a sudden change around the break radius, the pitch angle for the low frequency modes varies more smoothly with radius, which is a behaviour similar to that of the cavity modes of the control model. For a Type~II disc, we find that the pitch angle tends to increase with $r$.

\begin{figure*}
	\includegraphics[width=1\linewidth]{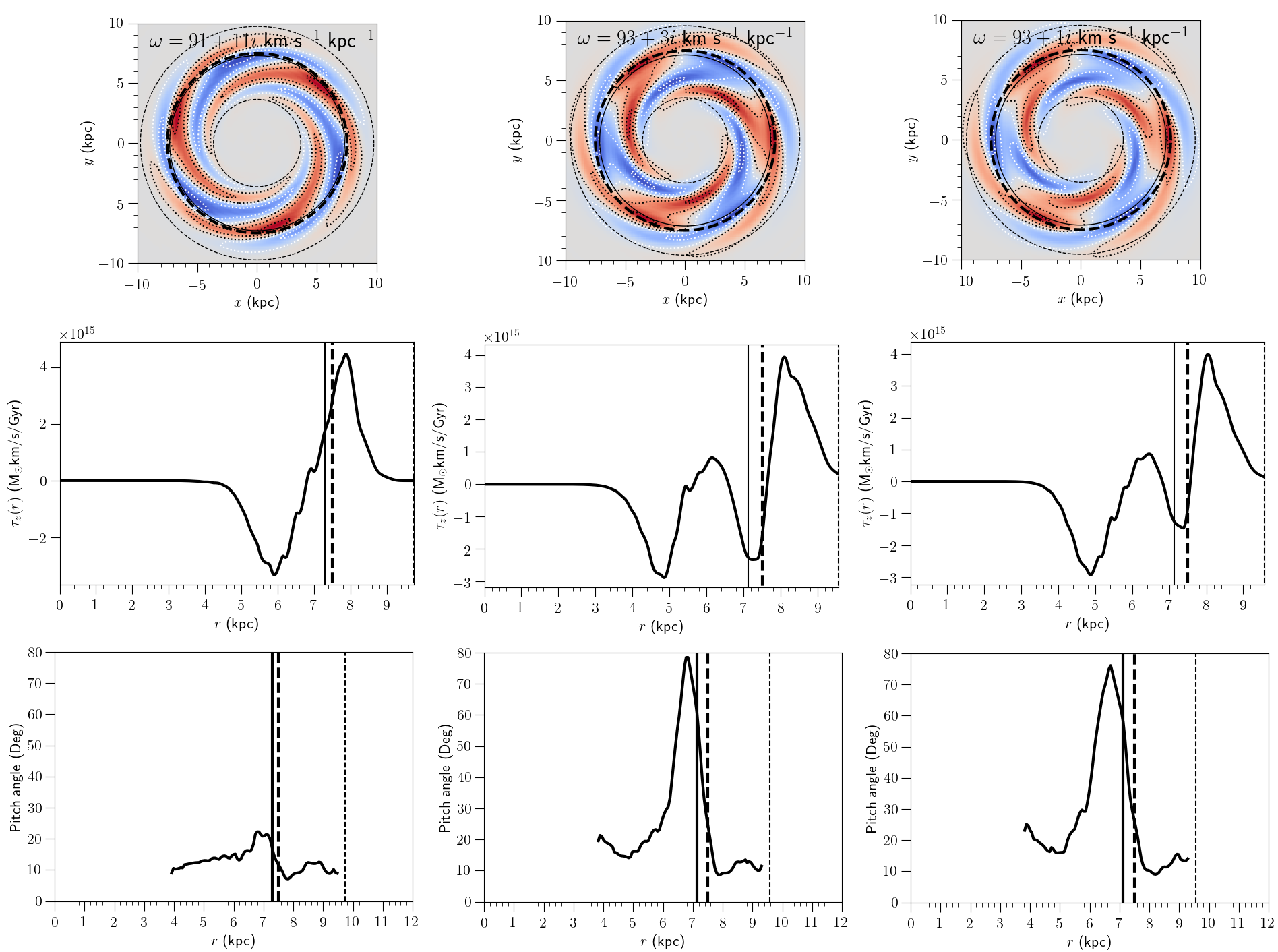}
    \caption{Top: density plots for the fast $m=3$ edge modes for a model with $f_{\rm br} = 0.5$ and $r_{\rm br}=7.5\kpc$. We test multiple versions of this model with different $Q$ profiles in the inner disc. From left to right $Q = 1.1,\, 1.5\, \text{and}\, 1.7$ in the inner disc. Middle: The torque profiles of the edge modes. Bottom: corresponding pitch-angle profiles for the edge modes. A cooler disc results in a density distribution which is more localized near the break, resulting in more localized torque profiles. The vertical full line marks the co-rotation radius of each mode, while the vertical dashed line indicates $\rbr = 7.5\kpc$.}
\label{fig:localised_modes}
\end{figure*}

\section{Varying the \texorpdfstring{$Q$}{\em{Q}}-parameter} \label{sec:cool_disc}

 All the models we have considered so far have identical Toomre-$Q$ profiles, with $Q$ hovering around $1.2$ inside \rbr\ (see Fig.~\ref{fig:mom}). Observational studies investigating disc stability have shown that nearby galaxies exhibit a wide range of values for the stellar $Q$-parameter $(Q \approx 1.4 - 3.2)$ across different morphological types \citep[e.g.][]{Romeo+2017, Aditya+2023}. We now investigate the impact of varying the $Q$-parameter in the inner disc has on the edge modes. We set up three models with $f_{\rm br} = 0.5$ and $r_{\rm br}=7.5\kpc$ and consider models with the $Q = 1.1$, $1.5$, and $1.7$\footnote{We also tested a model with $Q = 2.1$, which resulted in the edge modes being lost -- as predicted by \citet{Toomre1989}.}. Since we find that the effect which varying $Q$ has on modes of different $m$ is comparable, in Fig.~\ref{fig:localised_modes}, we show the resulting density distributions (top), torque (middle) and pitch angle profiles (bottom) for the fast (inner) $m = 3$ modes only, which are representative of the general trends we find. From left to right, we show inner-disc $Q = 1.1,\, 1.5\, \text{and}\, 1.7$. As with the previous models, $Q$ outside \rbr\ is not held fixed, but increases with $r$, which is a consequence of the decreased surface density. Changing $Q$ has only a small effect on the pattern speed but substantially suppresses their growth rates as $Q$ rises (the complex frequencies for the edge modes are shown in the top row of Fig.~\ref{fig:localised_modes}). This reduced growth rate is due to the less vigorous supporting response to the wake produced by an initial perturbation. We list the complex frequencies along with the resonant radii of the edge modes in Table.~\ref{table:3}.

\begin{table}
\centering
    \begin{tabular}{ ccccc }
        \multicolumn{5}{c}{$m=3$ edge mode frequencies for $r_{\mathrm{br}} = 7.5\kpc$ \& $f_{\mathrm{br}} = 0.5$} \\
        \hline
        $Q$ & $\Re({\omega})$ (km/s/kpc) & $\Im({\omega})$ (km/s/kpc) & CR (kpc)& OLR (kpc)\\
        \hline
        1.1 & 91.253& 11.283 & 7.305 & {\bf 9.746}\\
        1.1 &  67.573& 9.2047 & {\bf 9.469} & 12.815\\
        1.5 &  93.264& 3.494& 7.138  & {\bf 9.573}\\
        1.5 &  64.373& 5.305 & {\bf 9.835} & 13.444 \\
        1.7 &  93.487 & 1.393 &  7.1211  &  {\bf 9.555} \\
        1.7 &  63.465 &3.469 & {\bf 9.945} & 13.630 \\
             
        \hline
    \end{tabular}
\caption{Complex frequencies, and CR and OLR radii of the $m=3$ edge modes of a model with a density break with $r_{\mathrm{br}}~=~7.5\kpc$ and $f_{\mathrm{br}}~=~0.5$. Here, we vary the $Q$-parameter in order to investigate the impact this has on the edge modes. As with Table.~\ref{table:2}, the CR and OLR of the two members of each mode pair which show hints of resonance overlap are highlighted in bold-face.}
\label{table:3}
\end{table}

Fig.~\ref{fig:localised_modes} also shows that, as $Q$ increases, the overall density distribution (top) of the edge mode changes substantially, with the pattern broadening radially and developing a slight leading extension near the CR. This change is also reflected in the torque and pitch angle profiles of the edge modes. The slight leading extension results in an increase in the backwards torque inside the CR which is not present in the cooler, $Q=1.1$, model. In a hotter disc, the part of the spiral inside the CR is more open, as can be seen by the sharp increase in the pitch angle. Like the break models in Sec.~\ref{sec:typeII}, the spiral tightens quickly after the CR, with pitch angle values smaller than in the inner disc. Therefore, the tightening of the spiral after CR is generic, regardless of the $Q$-parameter. However, high $Q$ results in a spiral which is more open before CR due to the leading extension. Conversely, in a cooler disc (left), the overall pattern is less broad in the inner disc due to the lack of a leading extension, suggesting that the influence of the spiral becomes more localized to the break itself as the disc is cooled. Since the epicyclic motions in a cooler disc have a smaller amplitude, the small wavelength wakes are less damped, and the overall wakes are less spread out, resulting in the narrowing density distribution.

\begin{figure}
	\includegraphics[width=1\linewidth]{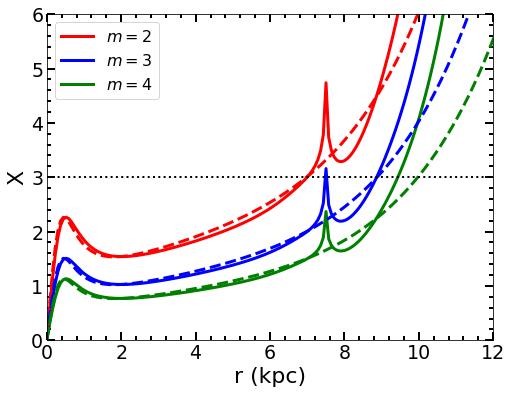}
    \caption{$X \equiv \lambda_{\rm \theta}/\lambda_{\rm crit}$ at the break as a function of radius for the control model (dashed) and a model with $r_{\rm br}=7.5~\kpc$ and $\fbr=0.5$ (solid). The horizontal dotted black line represents Toomre's condition for edge modes, $X \leq 3$. }
\label{fig:x_cond}
\end{figure}

\begin{figure}
	\includegraphics[width=\linewidth]{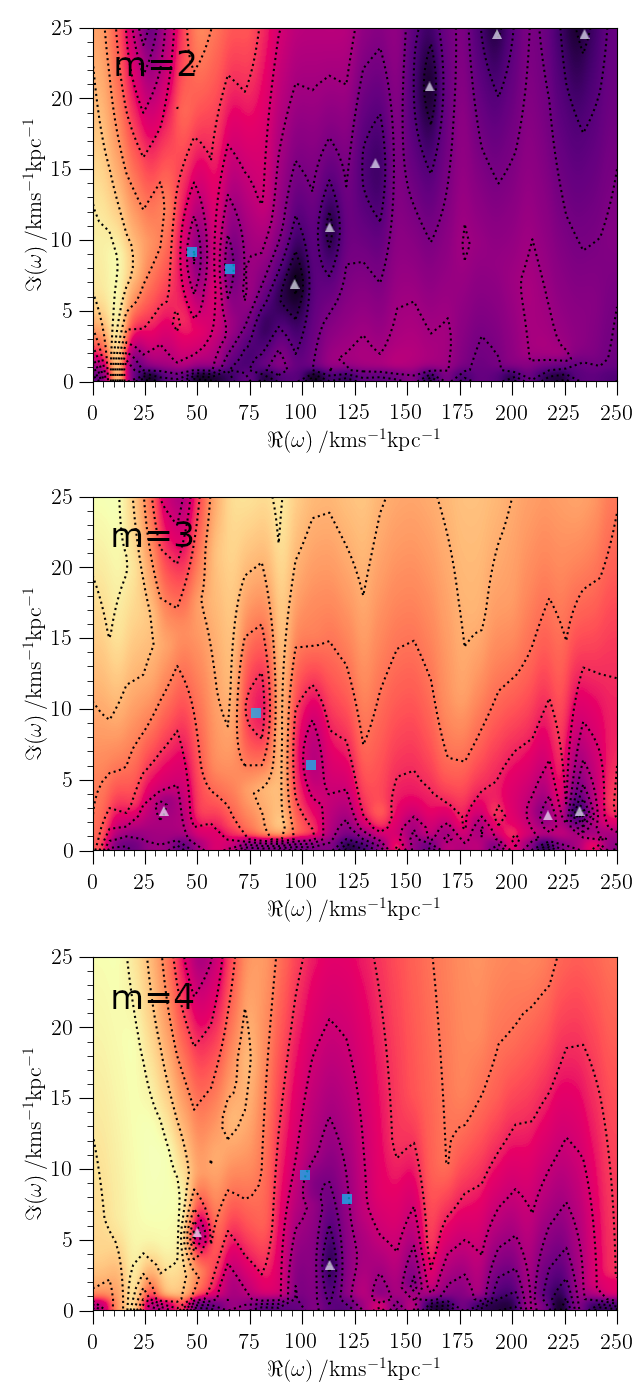}
    \caption{The $m=2,3,4$ complex frequency spectra for a model with a Type~III (up-bending) density profile (\fbr\ $= 2.0$) and \rbr\ $= 7.5$. Similar to models with Type~II breaks, Type~III breaks drive new modes which are not present in the control model.}
\label{fig:type_III_spec}
\end{figure}

\section{Edge Mode multiplicities}\label{sec:mult}

Throughout our analysis, we find that for the majority of the models we explored in Sec.~\ref{sec:typeII}, the $m = 2$ modes are the slowest growing of the fast $m=2-4$ modes. In Fig.~\ref{fig:rbr}, which shows the growth rates of the edge modes as a function of \rbr\ (left panel), the $m=2$ modes (red markers) consistently have the lowest growth rate until $r_{\rm br}=8~\kpc$, after which the growth rate for all $m$ become quite similar. The right panel of Fig.~\ref{fig:rbr},  which plots the edge mode growth rates as a function of \fbr, shows similar trends: at small values of \fbr, the $m=2$ modes grow significantly slower than the $m=3$ and $4$ ones, but as the break becomes weaker (for $f_{\rm br} > 0.75$), the fast edge modes of all $m$ have comparable growth rates. 

This behaviour can be understood qualitatively from \citet{Toomre1989}'s conditions for a disc to support edge modes. His first condition requires that the disc must be massive and cool enough to respond to disturbances in the phase space density, with $Q \leq 2$ and $X \equiv \lambda_{\rm \theta}/\lambda_{\rm crit} \leq 3$, where $ \lambda_{\rm \theta} = 2\pi r/m$ and  $\lambda_{\rm crit} \equiv 4\pi^{2}G\Sigma/\kappa^{2}$. The second condition is that the radial region over which the disc undergoes its most abrupt change must be smaller than one-fourth of the axisymmetric stability length, $\lambda_{\rm crit}$. The models employed in this study satisfy $Q \leq 2$ and since the breaks are abrupt, the latter condition is also satisfied. This leaves the condition $X \leq 3$ to be tested.

Fig.~\ref{fig:x_cond} shows how $X$ changes with radius. Here, the red, blue and green curves reflect the value of $X$ for $m=2$, $3$, and $4$, respectively. The dashed curves show the results for the control model, while the solid lines show $X$ for a truncated model with $\fbr=0.5$ and $\rbr=7.5\kpc$. At the location of the break, the curve for $m=2$ clearly exceeds $X=3$, while $m=3$ is just over this value. On the other hand, $X<3$ for $m=4$ for most of the disc. This suggests that the higher multiplicities will benefit more from amplification than waves with lower $m$. This could explain the trends observed in the previous section, where the higher-$m$ modes are more rapidly growing. However, we stress that while the preferred multiplicity is model specific -- depending on disc mass -- we find that the overall characteristics of the edge modes we observe are very similar, regardless of $m$.

\section{Edge modes in Type~III discs}\label{sec:type_III}

Thus far we have only considered modes in Type~II (i.e. down-bending, with $\fbr<1$) discs. Now we extend our study to Type~III (i.e. up-bending, with $\fbr>1$) discs.

\begin{figure*}
	\includegraphics[width=\linewidth]{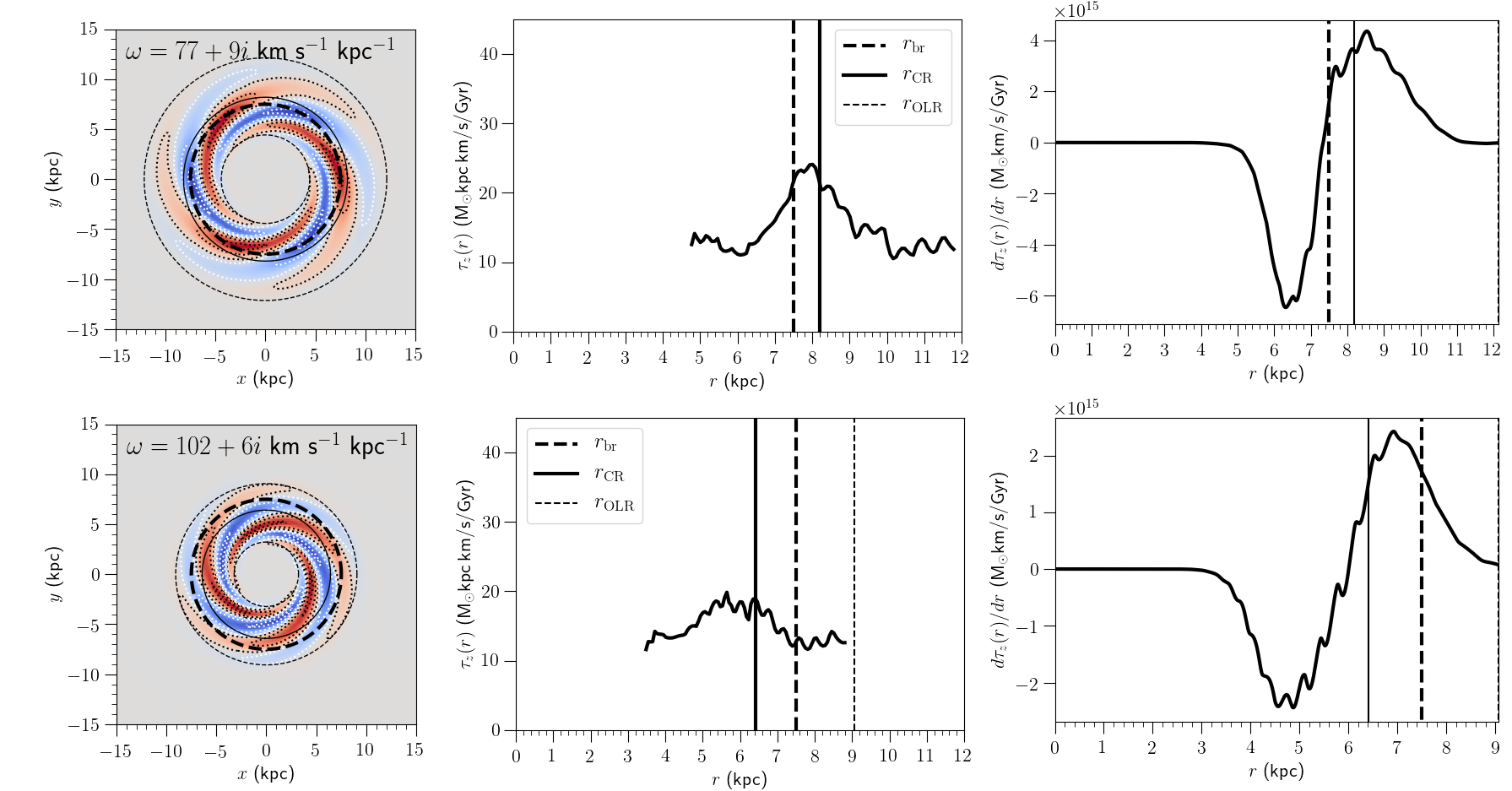}
    \caption{The surface density distribution (left), pitch angle profile (middle) and torque profile (right) for the slow outer (top) and fast inner (bottom) $m=3$ mode in a Type~III disc model with $r_{\rm br}=7.5 \kpc$ and $f_{\rm br} = 2.0$. The vertical full line marks the mode's CR radius; the dashed line indicates the location of the break.}
    \label{fig:typeIII}
\end{figure*}

In Fig.~\ref{fig:type_III_spec}, we show the $m=2$, $3$, and $4$ eigenmode spectra for a model with \rbr\ $= 7.5$ \kpc\, and \fbr\ $= 2.0$. Clearly, a Type~III break also has a destabilizing effect on the disc, leading to the growth of a number of new modes which are not present in the control model. As for the Type~II breaks, the edge modes also occur in pairs. In addition, the overall shape and torque profiles of the modes are qualitatively similar to the Type~II case (see left and right columns of Fig.~\ref{fig:typeIII}). 

However, in the case of Type~III breaks, we find that it is the outer (slow) mode which is consistently the faster growing of the edge mode pair, rather than the inner (fast) mode. This is opposite to what we observed in the Type~II cases. This behaviour is also demonstrated in Fig.~\ref{fig:rbr} (right panel), which shows the growth rate, $\Im({\omega})$, as a function of break strength, \fbr: for all $m$, the fast (inner) edge modes (triangles) have larger growth rates than the slow (outer) edge modes (circles) when $\fbr <1$. However, the trend is reversed when $\fbr>1$. Fig.~\ref{fig:rbr} also shows how the growth rates of the edge modes increase as the break gets stronger.

When compared with a Type~II mode, the Type~III edge mode pair also show opposite behaviour in terms of their pitch angle variation. In the middle column of  Fig.~\ref{fig:typeIII}, we show the pitch angle variation for the slow (upper panel) and fast (lower panel) edge modes. In a Type~III disc, it is the slow outer edge mode which shows an abrupt change in the pitch angle around the break region, rather than the fast inner mode. Conversely, the fast inner edge mode in a Type~III disc shows smoother variation in the pitch angle.

\begin{figure*}
	\includegraphics[width=\linewidth]{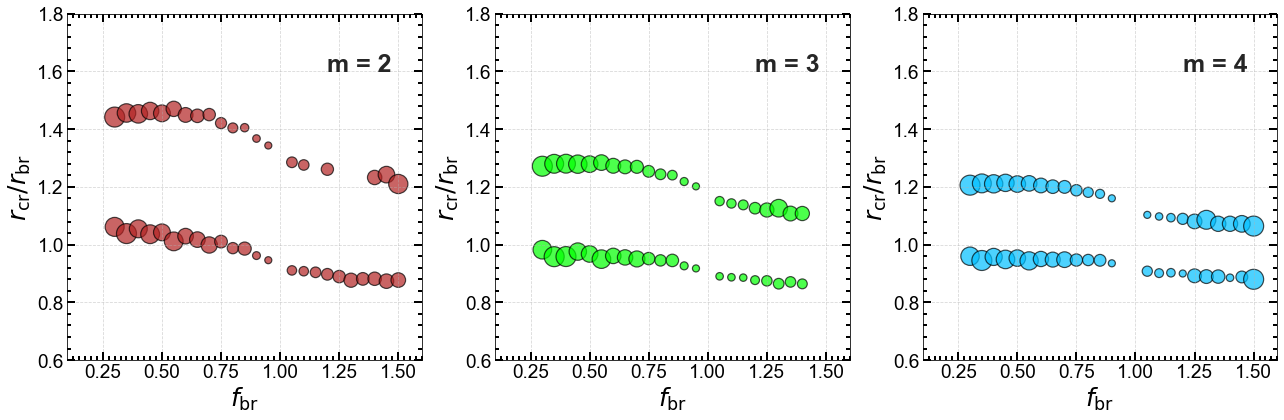}
    \caption{ The CR radius, normalized by \rbr, versus the break strength, \fbr. Here, $f_{\rm br} <1$ and $f_{\rm br} >1$ represent Type~II and Type~III profiles respectively, with the $f_{\rm br} = 1$ case being a single exponential disc model. The separate columns represent modes of different multiplicities. The size of each marker reflects the growth rate of the modes, with faster growing modes having larger markers.}
    \label{fig:break_strength2}
\end{figure*}

The higher growth rate for the slow outer mode is likely a result of the larger surface mass density in the outer disc of Type~III galaxies, allowing the outer edge mode to be more supported by the surrounding disc. However, when transitioning from a Type~II to a Type~III disc, the pattern speeds of the edge modes also shift, such that the CR of the slow outer mode is closer to the break. On the other hand, the CR of the fast inner mode shifts away from the break due to the increase in pattern speed. To demonstrate this, in Fig.~\ref{fig:break_strength2} we plot the ratio of the CR radius, $r_{\rm CR}$, and \rbr\ versus the break strength, \fbr\ . The size of the markers reflect the growth rate of the modes. When ($\fbr<1$), the fast inner mode has $r_{\rm CR}/\rbr \approx 1$. However, when $\fbr>1$ this ratio decreases for both edge modes, and results in the CR of the slow outer mode shifting towards the break. This could possibly explain the change in behaviour for the pitch angle for the edge modes. The edge mode with its CR closer to the break develops a more abrupt change in the pitch angle, while the other edge mode shows smoother pitch angle variation.

\section{Discussion}\label{sec:disc}

\subsection{Physical origin of edge modes} \label{sec:physical_origin}

In their study of discs with a number of ring-like features in the angular momentum distribution (such as grooves and ridges), \citet{Sellwood+1991} describe how such abrupt features induce sharp changes to the disc's impedance, which is given by the ratio of a pattern's perturbing gravitational force to the stellar displacement velocity it causes.

As such, the impedance is a measure of how strongly the stellar disc resists motion when subjected to a perturbing gravitational force. As is well known from wave theory, propagating waves are partially reflected at locations where the impedance changes abruptly. An abrupt change of the slope of the density profile, such as at the breaks of Type~II and Type~III discs, likewise induces a sharp change of the disc's impedance, and can, therefore, be expected to partially reflect travelling density waves. The same argument applies to an abrupt jump of the Toomre $Q$ stability parameter. As an aside, in Appendix~\ref{phase_space} we briefly explore the impact which a sharp change in the $Q$-profile has on the eigenmode spectrum of the disc. In agreement with \citet{Sellwood+1991}, we find that introducing jumps or dips in the $Q$-profile in a pure exponential disc also destabilizes modes. 

\citet{Binney2020} used linear theory to follow the evolution of the self-gravitating response of the shearing sheet to an imposed perturbation, such as a nearly axisymmetric density depression, or groove, that corotates with the sheet. He found that such a tilted groove `emits' density waves radially inwards and outwards. The more nearly axisymmetric the groove, the more rapidly these waves travel radially and the slower they decay. This appears to be a rather generic feature:~more general localized perturbations also trigger wave packets that travel away from their corotation radius and that have a decay time that declines as the perturbation is more axially symmetric. If these travelling waves lack a very strongly absorbing inner Lindblad resonance, the inward travelling waves can reflect at the centre, to form a resonant cavity between the groove and the galaxy centre, both of which reflect trailing into leading waves that can grow by swing amplification and set up a global, growing pattern.

It stands to reason that a similar evolution will unfold in the case of a density break, which, after all, can be regarded as a one-sided groove. This then offers a tentative explanation for the edge modes, with their co-rotation radius coincident with the break radius.

\subsection{Consequences for galaxy evolution in discs with a density break}
\label{sec:discussion}

The linear stability analysis we have carried out reveals that edge modes are excited by breaks. In Fig.~\ref{fig:break_strength2}, we plot the ratio of the CR radius, $r_{\rm CR}$, and \rbr\ versus the break strength, \fbr. The separate columns represent modes of different multiplicities, while the size of each marker reflects the growth rate of the modes, with faster growing modes having larger markers. Fig.~\ref{fig:break_strength2} shows that, in the case of Type~II profiles, the fast edge mode consistently has $r_{\rm CR}/r_{\rm br} \approx 1$ for all values of $m$.

In Sections~\ref{sec:torque&pitch} and \ref{sec:type_III}, we have explored the torque profiles generated by edge-modes and showed they generally increase the angular momentum of disc material in the broad vicinity of the break, pushing it outwards. This has interesting implications for the growth and evolution of disc galaxies. While a number of processes are responsible for the growth of a disc, observations support the idea that disc evolution is inside-out \citep{Vanderwel+2014,Rodriguez+2017}, including in the Milky Way \citep{Bovy+2012,Frankel+2019}. Disc breaks have also been observed in high redshift galaxies \citep{Perez2004}, with observational studies showing that the break radius increases with cosmic time \citep{Trujillo+2005,Azzollini+2008b}. 

Given the possible connection between disc breaks and galaxy growth, combined with the vigorously growing edge modes in cases where the break is strong enough, we envisage two possible scenarios for the growth of the outer disc: a Type~II disc break in a young disc -- seeded by a drop in the star-formation rate (SFR)\footnote{The drop in SFR can have various causes, including a drop in gas surface density, an extragalactic radiation field or the presence of a warp.} -- drives edge modes. A number of edge modes, with their CR straddling the disc break, therefore develop. These spiral modes drive radial migration across the break, adding mass to the outer disc and weakening the break in the process. As the star-forming disc grows due to the infall of fresh gas, the break is pushed further out, dragging along with it the edge modes. Consequently, as the edge-mode CR moves outward with the break, it is continuously driving radial migration across the break as the disc evolves, which changes the stellar demographics of the outer disc. 

Alternatively, the location of the break is primarily driven by the internal evolution of the disc, and not by details of the gas infall. In this view, the break itself also plays a role in moving the gas outward, leading to eventual star formation and inside-out growth. This -- coupled with stars being driven across the break due to the edge mode -- would imply that while the break is responsible for driving edge modes, the edge modes themselves erase the break from its current location and push it outwards. This scenario would have implications for the rate at which discs grow: as the break is pushed outward into regions of lower density, the surrounding response of the disc is diminished, resulting in weaker edge modes. Therefore, disc growth should slow down over time. Our {\sc pystab} analysis does not provide any insight into which of these scenarios is more likely. 

 The scenarios described above are consistent with Type~II breaks, however, the evolution of Type~III breaks do fit into either of these scenarios. In Sec.~\ref{fig:typeIII}, we showed that like the Type~II case, a Type~III disc will have an edge mode with its CR close to the break, resulting in mostly negative torque inside the break and positive torque outside. Stars immediately inside the break will gain angular momentum and get pushed into the outer disc. This will add mass to the outer disc, strengthening the break and driving more vigorous edge modes, which, in turn, drive more material into the outer disc. This constitutes a runaway cycle which continuously strengthens the Type~III break. However, the transfer of angular momentum also pushes the stellar break outward into regions of lower density, leading to more weakly growing edge modes and damping the runaway process. The fact that such a runaway process does not take place is supported by observations, which show that Type~III discs do not comprise the majority of galaxies in the local universe \citep[e.g.][]{Pohlen+2006,Erwin+2008}. Compared with Type~II breaks, observations of Type~III breaks also find that they are found at larger galactocentric radii; where the disc density is relatively low \citep[e.g.][]{Pohlen+2006, 2016A&A...596A..25L}. Still, our results based on linear theory do not capture any possible non-linear behaviour and understanding the evolution of Type~III breaks fully will require the analysis of detailed of $N$-body simulations.  

While the model employed to carry out this work is relatively simple and neglects the influence of gas, the star-forming hydrodynamical simulation analysed by \citet{Roskar+2012} also revealed a spiral with $r_{\rm CR}/r_{\rm br} \approx 1$ in their Type~II disc (see their Fig.~5), suggesting that edge modes also form in more realistically evolving systems. 

Simulations have also predicted that breaks in the density profile should also coincide with an upturn in the mean age profile, with the increase in mean age being interpreted as a signature of radial migration \citep{Roskar+2008a}. Indirect evidence of radial migration through age upturns has also been found in a number of observational studies \citep[e.g.][]{Bakos+2008,Yoachim+2010,Yoachim+2012,Radburn+2012}.

As we argued based on the torque profiles of the edge modes in Type~II profiles (cf. Sections \ref{sec:torque&pitch} and \ref{sec:type_III}), the expected effect of the secular evolution driven by these edge modes is that the break will become weaker, and less effective at causing or supporting edge modes. Hence, we can generally regard edge modes in stellar discs as an example of negative feedback in which a response of a physical system counteracts its cause.

\subsection{Observational signature of edge mode spirals}

One of the key parameters characterizing spirals is their pitch angle. Observational studies have found possible correlations between pitch angles and a number of structural parameters \citep[e.g.][]{Kennicutt+1982, Michikoshi+2014, Kendall+2015,Hart+2017,Diaz+2019,Font+2019,Yu+2019, Yu+2020}.

In Sec.~\ref{sec:torque&pitch}, we demonstrated how cavity and edge modes exhibit different behaviour in the radial variation of their pitch angle. While the pitch angle for cavity modes varies more gradually and over a larger radius, the changes in pitch angle for the edge mode with its CR close to the break is more abrupt, and occurs over a narrower region immediately around the break. In Sec.~\ref{sec:cool_disc}, we also showed that increasing the $Q$-parameter changes the overall shape of the edge mode, leading to an even larger change in the pitch angle around the break region.

In their study of 50 unbarred (or weakly barred) grand design spiral galaxies selected from SDSS, \citet{Savchenko+2013} found that in general spiral arms do not have constant pitch angles, but show a wide range of radial variation. For the majority of their sample, the pitch angle decreases with radius, but also observe cases where the pitch angle increases. \citet{Savchenko+2013} find differences in the pitch angle variation in galaxies with Type~II or Type~III breaks, with Type~III discs showing more pitch angle variation than Type~II discs (see their Fig 11). However, the driving mechanism of the spirals explored in this study is not known. Therefore, a direct comparison is not possible.

The abrupt change in the pitch angle for the edge mode we find in our work may provide a way of distinguishing it from other spiral modes in a disc. Therefore, this provides incentive for future observational studies to revisit this subject. 

\subsection{Resonant coupling}\label{sec:coupling}

In Sec.~\ref{sec:indentify_mode}, we demonstrated that, in a Type~II disc, the edge modes driven by breaks generally occur in pairs. Given that this study was carried out purely in the framework of linear theory, all the modes in this study grow independently and do not interact. However, an interesting feature which emerges from our analysis is that, for a given multiplicity, the OLR of the higher frequency mode is in close proximity to the CR of the lower frequency mode. In Table~\ref{table:2} we list the CR and OLR radii for the edge modes in a model with a Type~II disc. For a given $m$, we highlight in bold the frequencies which are in close proximity. The $m=2$ modes show the weakest potential coupling, with roughly $700\pc$ separating the OLR of the fast mode from the CR of the slower mode. However, coupling seems more likely for the $m=3$ and $m=4$ edge modes, with their respective resonance being separated by $100\pc$ or less. 

\begin{figure}
	\includegraphics[width=1\linewidth]{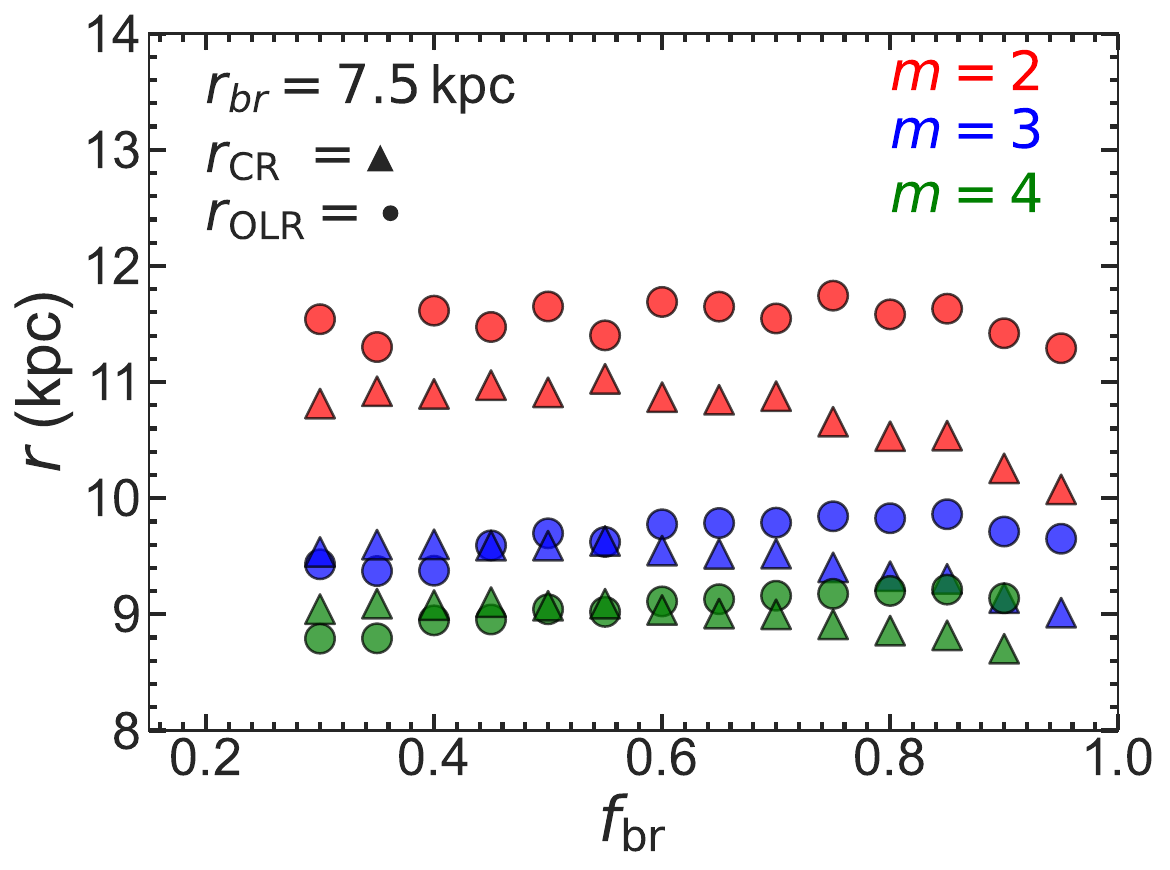}
    \caption{The radii of the fast mode OLR (circles) and slow mode CR (triangles) against the break strength, \fbr, for a model with $r_{\mathrm br}=7.5 \kpc$. Different colours represent different values of the multiplicity $m$. A stronger break results in closer resonances. As the break weakens, the relative distance between the resonances increases.}
\label{fig:res_overlap}
\end{figure}

In Fig.~\ref{fig:res_overlap}, we plot the OLR (circles) of the high frequency mode  and the CR (triangles) of the low frequency mode versus the break strength, \fbr. We do this for a model with $r_{\mathrm br}=7.5 \kpc$, and show the results for $m=2$ (red), $3$ (blue), and $4$ (green). In general, we find that a strong break leads to closer resonances. In the case of $m =3$ and $m=4$, the two resonances are, on average, separated by only $0.07 \kpc$ when $f_{\rm br}<0.6$, whereas the $m = 2$ resonances are separated by $0.6 \kpc$ on average. On the other hand, as the break becomes weaker ($f_{\rm br} \geq 0.6$), the resonances start to separate, with the OLR and CR being separated, on average, by $1.21$, $0.63$, and $0.43 \kpc$ for $m =2$, $3$, and $4$ respectively. Since the resonances approach each other in a region well outside the disc break itself ($r_{\mathrm br}=7.5 \kpc$), this permits stars to migrate even further into the disc outskirts.

In Sec.~\ref{sec:cool_disc} we also investigated the impact which changing the disc $Q$ profile has on the edge modes. We find that as the disc becomes hotter, the separation between the edge mode pair resonances increases. For example, when $Q=1.7$, the separation between the fast mode OLR and slow mode CR is roughly $0.4\kpc$, while the separation in a $Q=1.1$ model is only $0.27\kpc$ (see Table.~\ref{table:3}). While the pattern speeds do not vary substantially, an increase in $Q$ leads to a decrease in the pattern speed of the slow outer edge mode, resulting in larger separation of the resonances. 

Therefore, a strong break leads not only to more vigorously growing edge modes, which redistribute material across their resonances, but our analysis also suggests that there is a higher probability of resonant spiral-spiral coupling, which further increases the efficiency of the redistribution of material across the outer disc \citep[e.g.][]{Minchev+2011}

\subsection{Caveats}

 An important caveat to this work is that it omits the influence gas and disc thickness, which impact the stability characteristics of the disc \citep[e.g.][]{Jog+1984,Elmegreen+2011,Romeo+2011, Shadmehri+2012, Hoffmann+2012, Romeo+2013}. \citet{Romeo+2013} carried out a multi-component stability analysis of galaxies from The HI Nearby Galaxy Survey (THINGS), and found that while stars largely dominate the disc stability beyond one scale-length, molecular gas also provides a significant contribution. Using the same survey, \citet{Hoffmann+2012} also showed that ISM turbulence has a significant effect across the disc, and makes the outer disc more prone to star-dominated instabilities. The effective $Q$-parameter is also impacted by disc thickness \citep{Elmegreen+2011,Romeo+2011}. In particular, \citet{Romeo+2011} find that disc thickness increases the $Q$-parameter by $\approx 20-50\%$.

This work also ignores the influence of a bar. While we do not expect that the presence of a bar will alter the stability characteristics of the disc, studies have found hints that bars may play a role in setting the break radius. In an observational study of $218$ nearby disc galaxies, \citet{Mateos+2013} found that, for galaxies more massive than $10^{10}~M_{\rm \odot}$, the distribution of the ratio of the break and bar radii has two peaks at $r_{\rm br}/r_{\rm bar}~\approx~2$ and $r_{\rm br}/r_{\rm bar}~\approx~3.5$. They link the first value with the bar OLR and the second with a resonant coupling between the bar and a spiral. 

Investigating the impact which bars, gas, and disc thickness have on our results requires the detailed dissection of $N$-body+SPH simulations, which is beyond the scope of this paper. Furthermore, it is also beyond the capabilities of \textsc{pystab}, which needs to start from an axially symmetric initial state within a razor-thin disc.

\subsection{Summary}\label{sec:conclusion}

By analysing the stability of disc galaxy models using linear perturbation theory, we have investigated the role that breaks play in driving edge modes in stellar discs. The main conclusions of this paper are the following:
\begin{enumerate}
\item Introducing a density break (either a Type~II or III) in the disc has a destabilizing effect which gives rise to a number of vigorously growing $m = 2$, $3$, and $4$ edge modes. These edge modes are absent from the control model lacking a disc break. 

\item Both a Type~II and Type~III break drive edge modes in pairs having different characteristics. In the Type~II model, the inner "fast" mode has its CR roughly coincident to the break, and has a higher growth rate. Additionally, it shows an abrupt increase in pitch angle at the break. In contrast, the "slow" outer mode -- with its CR further from the break -- shows a smoother variation of pitch angle over a large radius. 

\item Transitioning from a Type~II to a Type~III disc causes the pattern speeds of the edge modes to increase. Consequently, it is the slow outer mode in a Type~III which has its CR closer to the break. This, combined with the increased mass in the outer disc, results in a higher growth rate for the "slow" mode. In addition, it is the "slow" mode which shows an abrupt change in the pitch angle around the break, with the "fast" mode showing a smoother variation.

\item Increasing the break radius results in a decrease in the growth rates of the edge modes. This is caused by the weaker torque of the disc in response to perturbations at the break.
\item Likewise, the break strength, in the sense of the change of slope at the break, plays an important role in producing edge modes. We find that in both Type~II and Type~III profiles, making the break more abrupt results in a substantial increase in the growth rates of the edge modes.

\item For every multiplicity we investigated, Type~II breaks result in coupled modes. The CR of the high-frequency member of the couple is almost coincident with the break radius. Moreover, its OLR is in close proximity to the CR of the low-frequency mode. This may lead to resonant coupling between these modes when they enter the non-linear regime.

\item The $Q$-parameter has a substantial effect on the pitch angle and resulting torque profiles of the spiral. Increasing $Q$ leads to a spiral which is more radially broad in the vicinity of the break, with a leading extension just inside the break that quickly tightens at the break. This results in a sudden increase (and decrease) in the pitch angle in this region.

\end{enumerate}

\section*{Acknowledgements}
 KF is partially funded by the the Tertiary Education Scholarships Scheme (TESS, Malta). KF wishes to thank C. F. P. Laporte for his valuable comments on this work. S.D.R. acknowledges support from grant Segal ANR-19-CE31-0017 of the French Agence Nationale de la Recherche. S.D.R. wishes to thank B. G. Elmegreen and F. Combes for their valuable comments on this work.

We thank A. Romeo for his constructive comments which helped to improve this paper.


\section*{Data availability}
The software and model data underlying this article are a mix of proprietary and available; some data may be shared upon reasonable request to K. F. (karl.fiteni.12@um.edu.mt).



\bibliographystyle{mnras}
\bibliography{refs} 



\appendix

\section{Linear stability analysis code}\label{pystab}

The dynamical properties of an axially symmetric collisionless disc are captured fully by the underlying gravitational potential $V_{0}(r)$, along with its phase space distribution function (DF), $F_{0}(E,J_{\theta})$. Here, $E$ represents the specific binding energy of a star and $J_{\theta}$ the specific angular momentum of a stellar orbit. \textsc{pystab} obtains the complex frequencies $\omega$ for which a spiral-shaped perturbation of the form
\begin{equation}
   \centering
   V_{\rm pert}(r,\theta,t) = V_{\rm pert}(r)e^{i(m\theta-\omega t)}
\end{equation}
constitutes an eigenmode with an infinitesimal amplitude. Here, $(r,\theta)$ are polar coordinates in the stellar disc, and $V_{\rm pert}(r)$ is a complex function which quantifies the amplitude and phase of the mode. Additionally, $m$ represents the multiplicity (or radial symmetry) of the mode, while $\Omega_p = \Re(\omega)/m$ and $\Im(\omega)$ represent its pattern speed and the growth rate. 

In the linear regime, any general perturbing potential can be expanded as a series of spirals, which may be analysed independently. \textsc{pystab} determines the response distribution function $f_{\rm resp}(r,\theta,v_{\rm r},v_{\rm \theta},t)$ produced by a perturbation $V_{\rm pert}(r,\theta,t)$ by solving the first-order collisionless Boltzmann equation,
\begin{equation}
    \left. \frac{D f_{\text{resp}}}{Dt} \right|_0
    = - \frac{\partial F_0}{\partial \vec{v}}.\vec{\nabla}V_{\text{pert}},
\end{equation}
where the left-hand side contains the time-derivative of the response distribution function along an unperturbed orbit. The formal solution to this equation is given by
\begin{equation}
    f_{\rm resp}(\vec{r},\vec{v},t) = -\frac{\partial F_0}{\partial \vec{v}}. \int_{-\infty}^t 
    \vec{\nabla}V_{\text{pert}}(\vec{r}(t),t') dt', \label{formsol}
\end{equation}
where it is tacitly assumed that the response disappears in the infinite past.

The response distribution function $f_{\rm resp}(r,\theta,v_{\rm r},v_{\rm \theta},t)$ produces a response density $\Sigma_{\rm resp}(r,\theta,t)$ of the form
\begin{equation}
   \centering
   \Sigma_{\rm resp}(r,\theta,t) = \int f_{\rm resp}(r,\theta,v_{\rm r}, v_{\rm \theta},t)dv_{\rm r}dv_{\rm \theta}.
\end{equation}
From this, the corresponding gravitational response is also obtained
\begin{equation}
   \centering
   V_{\rm resp}(\bar{r}) = G \int \Sigma_{\rm resp}(\bar r')\psi(|\bar r - \bar r'|)d^{2}\bar r' ,
\end{equation}
where $\psi$ is the inter-particle interaction potential, which in a Newtonian framework, is given by
\begin{equation}
   \centering
 \psi(|\bar r - \bar r'|) = \frac{1}{|\bar r - \bar r'|} .
\end{equation}
and is used for all interactions. The eigenmodes of the disc are determined by the condition
\begin{equation}
   \centering
 V_{\rm pert}(r,\theta,t) \equiv V_{\rm resp}(r,\theta, t),
\end{equation}
which \textsc{pystab} determines via a matrix method \citep{Kalnajs1977,Vauterin+1996}. In order to locate the eigenmodes, $V_{\rm pert}$ is expanded as a series basis potentials, $V_{\rm l}$. The response to each individual basis potential, $V_{\rm l, resp}$, can again be expressed as a series:
\begin{equation}
   \centering
V_{\rm l, resp} = \sum_k C_{\rm lk} V_{\rm k}.
\end{equation}

\textsc{pystab} exploits the fact that the $C$ matrix, which is $\omega$-dependent, will have a unity eigenvalue if the perturbation is an eigenmode \citep{Vauterin+1996}. The amplitude at $t = -\infty$ is assumed to be zero. Therefore, only eigenmodes with positive growth rates ($\Im(\omega)>0$) are considered.

A number of technical parameters were employed in order to set up the model. Firstly, the number of orbits on which phase space is sampled is given by $n_{\rm orbit}(n_{\rm orbit}+1)/2$, where $n_{\rm orbit} = 600$. Secondly, the number $n_{\rm Fourier}$, which reflects the number of Fourier components in which the periodic part of the perturbing potential is expanded. We took this value to be $n_{\rm Fourier} = 80$. Additionally, we used a total of 60 potential-density pairs (PDPs) for the expansion for the radial part of the perturbing potential and density. We use PDP densities with compact radial support which cover the relevant part of the stellar disc and are evenly spaced on a logarithmic scale, resulting in the highest resolution in the inner regions of the disc. Their radial widths are automatically chosen such that consecutive basis functions are unresolved and can be used to represent any smooth radial function. The corresponding PDP potentials are computed numerically using equation (A3).

\section{Breaks in phase space}\label{phase_space}

As an aside, we note that, from the impedance argument, one expects that an initial phase-space perturbation need not have a corresponding density signature to excite waves. Indeed, \citet{Sellwood+1991} find that when a groove is introduced in the angular momentum space of a warm disc, the disc is still destabilized even when the groove is masked in the density distribution by the epicyclic motion of the stars. This can be seen from the formal solution of the linearized Boltzmann equation, given by Eqn.~\ref{formsol}. The response distribution function depends on the velocity gradient of the distribution function of the unperturbed, zeroth-order galaxy model. If the latter is altered such that its velocity gradient changes while its corresponding density distribution remains the same, the corresponding spectrum of eigenmodes will be affected and eigenmodes may pop up where previously there were none \citep{2019MNRAS.484.3198D}. 

To investigate whether we can induce new modes in our fiducial model without changing its density profile, we run a number of models which include either a sudden rise or dip in the $Q$ profile at $r=7.5~\kpc$ while keeping its exponentially declining radial density profile fixed. We model the Q-profile using the following function, which introduces the abrupt feature at \rbr:
\begin{equation}\label{eq:Q_prof}
     Q(r) = \pm a\arctan(b(r-\rbr))+c.
 \end{equation}
Here, $a$ controls how large the change in $Q$ is at \rbr, while $b$ reflects the abruptness of the change. Lastly, $c$ is a simple offset. In agreement with \citet{Sellwood+1991}, we find that any sudden change in the $Q$-profile results in a number of new modes for $m=2,\ 3$, and $4$. In Sec.~\ref{sec:typeII}, we showed how a density break gives rise to an edge mode with its CR straddling the break. Similarly, here we also find that, for all $m$, the fastest growing edge mode has its CR close to $7.5\kpc$ - the radius where the change in $Q$ is introduced. We also find that the range of pattern speeds and growth rates are in the same range as those found in the density break models. Unsurprisingly, the growth rates are mainly dependent on the size of the change in $Q$ (constant $a$ in Eq.~\ref{eq:Q_prof}). Analysing these modes further, we find their density distribution and torque profile to be similar to those of the edge modes in the previous models. 

We conclude that sharp features in phase space are able to drive modes, irrespective of whether they have a corresponding spatial-density signature. This supports the argument that travelling waves being reflected off abrupt impedance jumps in the underlying disc are causing the break-related modes we find (be that density breaks or $Q$ breaks).

\bsp	
\label{lastpage}
\end{document}